\documentclass[twocolumn,prd,showpacs,preprintnumbers,unsortedaddress,superscriptaddress,amsmath,amssymb]{revtex4-1}
\usepackage{graphicx,slashed,bbold,subfigure}
\usepackage[latin1]{inputenc}
\usepackage{placeins}

\begin{document}
\title{Nonuniform phases in a three-flavor Nambu--Jona-Lasinio model}

\author{J. Moreira }
\affiliation{Centro de F\'{i}sica Computacional, Departamento de F\'{i}sica da Universidade de Coimbra, 3004-516 Coimbra, Portugal}
\author{B. Hiller}
\affiliation{Centro de F\'{i}sica Computacional, Departamento de F\'{i}sica da Universidade de Coimbra, 3004-516 Coimbra, Portugal}
\author{W. Broniowski}
\affiliation{The H. Niewodnicza\'nski Institute of Nuclear Physics, Polish Academy of Sciences, PL-31342 Krak\'ow, Poland
%\\
}
\affiliation{
Institute of Physics, Jan Kochanowski University, PL-25406~Kielce, Poland}
\author{A. A. Osipov}
\affiliation{Centro de F\'{i}sica Computacional, Departamento de F\'{i}sica da Universidade de Coimbra, 3004-516 Coimbra, Portugal}
\affiliation{on leave from Dzhelepov Laboratory of Nuclear Problems, JINR 141980 Dubna, Russia}
\author{A. H. Blin}
\affiliation{Centro de F\'{i}sica Computacional, Departamento de F\'{i}sica da Universidade de Coimbra, 3004-516 Coimbra, Portugal}

\begin{abstract}
It is shown that  flavor mixing of the strange and light quarks  allows for existence of a much larger baryonic chemical potential window for the formation of a stable dual chiral-wave state as compared to the well known two flavor case. In addition, strangeness catalyzes the occurrence of a new branch of non-homogeneous solutions at moderate densities. 
% and moderate values of the chiral density wave momentum $q$. 
This case study is addressed at zero temperature within the SU(3) flavor Nambu-Jona-Lasinio model with the 't Hooft determinantal flavor mixing interaction. The modulation of the chiral condensates in the light quark sector is taken to be one-dimensional, while strangeness is embedded as a homogeneous condensate in the spontaneously broken phase of chiral symmetry. A finite current quark mass for the strange quark is incorporated, while the up and down current masses are set to zero. In that case the modulation considered provides an exact analytic solution for the system. Despite the simplicity of the \textit{ansatz}, the emerging phase diagram 
% in the chemical potential versus mixing strength 
displays a very rich structure.
\end{abstract}

%Keywords: QCD, Explicit symmetry breaking, Effective theory, Nambu--Jona-Lasinio model, phase transitions, quark matter, inohomogenous phases, pion condensation.

\maketitle

%%%%%%%%%%%%%%%%%%%%%%%%%%%%%%%%%%%%%%%%%%%%%%%%%%%%%%%%%%%%%%%%%%%%%%%%%%%%
%                        main text
%%%%%%%%%%%%%%%%%%%%%%%%%%%%%%%%%%%%%%%%%%%%%%%%%%%%%%%%%%%%%%%%%%%%%%%%
\section{Introduction}
In the region of the low temperature ($<200$~MeV) and moderately high baryon chemical potential (that is, baryon densities in the range up to a few nuclear saturation densities) in the phase diagram of strongly interacting matter (for a recent review see, e.g., \cite{Fukushima:2010bq} and references therein), the use of effective models is of particular importance. The popularity of such studies stems from the known difficulties of more fundamental approaches such as the first-principle lattice QCD. The Nambu-Jona-Lasinio (NJL) model~\cite{Nambu:1961tp, Nambu:1961fr, Vaks:Larkin} and its extensions is widely regarded as a basic tool, as it shares with QCD its global symmetries and incorporates a mechanism for dynamical chiral symmetry breaking. $U(1)$ axial symmetry, which is not observed in nature, can be explicitly broken in the model by including the 't Hooft determinant in the Lagrangian (NJLH)~\cite{'tHooft:1976fv, Bernard:1987gw, Bernard:1987sg, Reinhardt:1988xu}. The confinement effects may be mimicked by the phenomenological introduction of the Polyakov loop~\cite{Fukushima:2003fw,Megias:2004hj,Ratti:2005jh,Roessner:2006xn}. In the applications to the dense media, the NJL model, carrying the quark degrees of freedom, may  hopefully be adequate at densities where nucleons melt out into constituent quarks. 

The possibility of the appearance of non-uniform phases in the QCD phase diagram has been proposed long ago (for a recent historical review see, for instance,~\cite{Broniowski:2011ef}). The effect follows from the fact that the pion's interaction with nucleons or quarks is attractive when the mean pion field carries a gradient, thus making nonuniformity favorable. At the same time the kinetic term suppresses the gradients, thus competing with the gradient term in a non-trivial dynamics. One should note that these ideas are borrowed from condensed matter~\cite{Larkin:1964zz, Fulde:1964zz}.

Following the $p$-wave pion condensation in nuclear matter proposed by Migdal in~\cite{Migdal:1971cu, Migdal:1973zm}, the subject has been studied by numerous authors. The generalization to  relativistic systems was considered in~\cite{Dautry:1979bk, Broniowski:1990dy, Nakano:2004cd}, the large-$N_c$ arguments were used for the large density and zero temperature case in~\cite{Deryagin:1992rw, Shuster:1999tn}, while the case of quarkyonic matter was explored in~\cite{Kojo:2009ha,Kojo:2010fe,Kojo:2011cn,Partyka:2010em,Carignano:2010ac}. Recently, the Dyson-Schwinger approach to the problem was investigated in~\cite{Muller:2013pya,Muller:2013tya}. The possibility of two dimensional modulations was studied in~\cite{Carignano:2011gr,Carignano:2012sx} within the NJL model. The embedding of solutions of the Gross-Neveu model allowed for analytic studies of the non-uniform states of quark matter in a model without the pion field~\cite{Nickel:2009ke,Nickel:2009wj,Buballa:2009ct}. The effects of the non-zero current quark masses in spatially inhomogeneous chiral condensates were studied perturbatively in~\cite{Maedan:2009yi}. This scenario is studied in nuclear matter using an extended linear sigma model in \cite{Heinz:2013hza}. The magnetic features of the nonuniform phase have been discussed in~\cite{Kotlorz:1994sk, Takahashi:2007qu, Frolov:2010wn, Basar:2010zd, Tatsumi:2011tu, Rabhi:2011mj,Ferrer:2007iw}. 
%%NEWrefs!!!
Inhomogeneous phases in isotopically asymetric dense quark matter are considered in \cite{Mu:2010zz, Ebert:2011rg,Gubina:2012wp}.

Methodologically, the chiral-density wave scenario is analogous to the spin-density wave scenario proposed in \cite{Overhauser:1962zz}. The same underlying particle-hole pairing mechanism has also been considered in the study of color-superconductivity \cite{Alford:2000ze,Bowers:2002xr,Casalbuoni:2005zp, Mannarelli:2006fy, Rajagopal:2006ig, Nickel:2008ng, Sedrakian:PhysRevD.80.074022,Ebert:2013dda}.

%At the Facility for Antiproton and Ion Research (FAIR) at GSI and the Nuclotron-based Ion Collider Facility (NICA) at JINR an experimental effort to gain a better understanding of strongly interacting matter in the low temperature/high chemical potential regime is going to be made. Possible tell-tale signals for the occurrence of a spatially inhomogeneous phase such as the production of opposite multipion jets, with individual pions having momenta of the order of several hundred MeV suggested in \cite{Sadzikowski:2000ap} can then be searched.

In this work we investigate the appearance of a phase with an one-dimensional spatial modulation in the chiral condensates in the three-flavor case with vanishing current masses for the $u$ and $d$ quarks but with a finite mass for the $s$ quark.

\section{The model}
Relaxation of the homogeneity constraint opens a whole new world of possibilities for the spatial modulation of the scalar and pseudoscalar chiral condensates. One simple \textit{ansatz}, which we will use in the present work, is the dual chiral-density wave suggested in \cite{Dautry:1979bk}. It corresponds 
to the following form of the light-quark ($u$, $d$) condensates:
\begin{align}
\langle\overline{\psi_l}\psi_l\rangle=\frac{h_l}{2} \mathrm{cos}(\boldsymbol{q}\cdot\boldsymbol{r}), \quad
\langle\overline{\psi_l} i \gamma_5\tau_3\psi_l\rangle=\frac{h_l}{2} \mathrm{sin}(\boldsymbol{q}\cdot\boldsymbol{r}), \label{eq:dn}
\end{align}
where $\tau_3$ corresponds to the Pauli matrix acting in isospin space. For the strange quark a uniform condensate background is considered. 
The ansatz results in the modification of the single-particle light-quark energy spectrum \cite{Dautry:1979bk}:
\begin{align}
\label{ESpectrum}
E^{\pm}&=\sqrt{M^2+p^2+\frac{q^2}{4}\pm\sqrt{\left(\boldsymbol{p}\cdot\boldsymbol{q}\right)^2+M^2 q^2}},
\end{align}
where $M$ is the dynamical mass, $\boldsymbol{p}$ denotes the momentum of the quark, 
and $\boldsymbol{q}$ is the wave number of Eq.~(\ref{eq:dn}). We chose the $z$-axis to coincide with $\boldsymbol{q}$. Note that the $E^-$ branch has a lower energy, thus its occupation is preferable. Quite remarkably, the ansatz (\ref{eq:dn}) and the corresponding quark orbitals with energies~(\ref{ESpectrum}) form a self-consistent
solution of the Euler-Lagrange equations. 

%%NEWmodellagrangian!!!
Our starting point is the model Lagrangian expressed in terms of the quark fields $\psi(\overline{\psi})$:
\begin{align}
\label{modellagrangian}
\mathcal{L}=&\mathcal{L}_D+\mathcal{L}_{NJL}+\mathcal{L}_{H}\nonumber\\
\mathcal{L}_D=&\overline{\psi}\left(\imath\gamma^\mu\partial_\mu-m\right)\psi\nonumber\\
\mathcal{L}_{NJL}=&\frac{G}{2}\left(\left(\overline{\psi}\lambda_a \psi\right)^2+\left(\overline{\psi}\imath\gamma^5\lambda_a \psi\right)^2\right)\nonumber\\
\mathcal{L}_{H}=&\kappa\left(\mathrm{det}\left(\overline{\psi}\frac{1-\gamma^5}{2} \psi\right)+\mathrm{det}\left(\overline{\psi}\frac{1+\gamma^5}{2} \psi\right)\right),
\end{align}
where $m$ corresponds to the current mass diagonal matrix, $\lambda_a$ are Gell-Mann flavor matrices and $\mathrm{det}$ is the flavor determinant.
%The chiral projectors are as usual represented by $P_{L(R)}$.

Using the techniques of Ref.~\cite{Osipov:2003xu}, the thermodynamic potential of the model in the mean field approximation is given by
% (note that the strange quark is not affected directly by the spatial modulation)
\begin{align}
\label{Omega}
\Omega=&V_{st}+\frac{N_c}{8\pi^2}\times\nonumber\\
&\left(J_{-1}(M_u,\mu_u,q)+J_{-1}(M_d,\mu_d,q)+J_{-1}(M_s,\mu_s,0)\right)\nonumber\\
V_{st}=&\frac{1}{16}\left.\left(4G\left(h_u^2+h_d^2+h_s^2\right)+\kappa h_u h_d h_s\right)\right|^{M_i}_0,
\end{align}
where $h_i$ ($i=u,d,s$) are twice the quark condensates. The integrals $J_{-1}$ stem from the fermionic path integral over the quark
bilinears which appear after bosonization, while $V_{st}$ corresponds to the stationary phase contribution to the integration over the auxiliary bosonic fields. From the value evaluated at the dynamical masses $M$, a subtraction of its value evaluated at $M=0$ is made \cite{Hiller:2008nu} (which is what is meant by the $|^M_0$ notation in the last line of Eq.~(\ref{Omega})).

Using a regularization kernel corresponding to two Pauli-Villars subtractions in the integrand \cite{Pauli:1949zm,Osipov:1985}, previously used for instance in \cite{Osipov:2006xa, Osipov:2007mk}, namely
%\begin{align}
$\rho\left(s\Lambda^2\right)=1-(1+s\Lambda^2)\mathrm{exp}(-s\Lambda^2)$,
%\end{align} 
%\cite{Hiller:2008nu,Moreira:2010bx}. 
%Due to the 4D symmetry breaking induced by the appearance of the spatial modulation we chose to operate the regularization in the component of the momentum perpendicular to $\overrightarrow{q}$:
%\begin{align}
%\hat{\rho}_{p_\perp}=& 1-(1-\Lambda^2\frac{\partial}{\partial
%p_\perp^2})e^{\Lambda^2\frac{\partial}{\partial p_\perp^2}}.
%\end{align}
the Dirac and Fermi sea contributions, $J^{vac}_{-1}$ and $J^{med}_{-1}$, can be written as
\begin{align}
J_{-1}=&J^{vac}_{-1}+J^{med}_{-1},\nonumber\\
J^{vac}_{-1}=&
\int\frac{\mathrm{d}^4 p_E}{(2\pi)^4}\int^\infty_0 \frac{\mathrm{d}s}{s}\rho\left(s\Lambda^2\right)8\pi^2e^{-s\left(p_ {0\,E}^2+p_\perp^2\right)}\nonumber\\
&\left.\left(e^{-s\left(\frac{q}{2}+\sqrt{M^2+p_z^2}\right)^2}+e^{-s\left(\frac{q}{2}-\sqrt{M^2+p_z^2}\right)^2}\right)\right|^{M,q}_{0,0},\nonumber\\
J^{med}_{-1} =&-\int\frac{\mathrm{d}^3p}{(2\pi)^3}8\pi^2T\left.\left(\mathcal{Z}^+_++\mathcal{Z}^+_-+\mathcal{Z}^-_++\mathcal{Z}^-_-\right)\right|^{M,q}_{0,0}\nonumber\\
              &+C(T,\mu),\nonumber\\
\mathcal{Z}^\pm_\pm =&\mathrm{log}\left(1+e^{-\frac{E^\pm\mp\mu}{T}}\right)-\mathrm{log}\left(1+e^{-\frac{E_\Lambda^\pm\mp\mu}{T}}\right)-\nonumber\\
&\frac{\Lambda^2}{2T E_\Lambda^\pm}\frac{e^{-\frac{E_\Lambda^\pm\mp\mu}{T}}}{1+e^{-\frac{E_\Lambda^\pm\mp\mu}{T}}},\nonumber\\
C(T,\mu)=&\int\frac{\mathrm{d}^3p}{(2\pi)^3}16\pi^2T~\nonumber\\
&\mathrm{log}\left(\left(1+e^{-\frac{|\boldsymbol{p}|-\mu}{T}}\right)\left(1+e^{-\frac{|\boldsymbol{p}|+\mu}{T}}\right)\right)
\end{align}
where $E_\Lambda^{\pm}=\sqrt{\left(E^{\pm}\right)^2+\Lambda^2}$. The $|^{M,q}_{0,0}$ notation refers to the subtraction of the same quantity evaluated for $M=0$ and $q=0$, which is done so as to set the zero of the potential to a uniform gas of massless quarks (it amounts to a subtraction of a constant). The superscript $\pm$ in the definition of $\mathcal{Z}$ refers to the energy branch, whereas the subscript refers to the sign in front of the chemical potential in the exponent. The $C(T,\mu)$ term is needed for thermodynamic consistency \cite{Hiller:2008nu}%,is set to counterbalance the first part of the $\mathcal{Z^\pm_\pm}$ terms in the zero mass subtractions
.

The minimization of the thermodynamical potential with respect to $M$ and $q$ has to be done self-consistently via solving the stationary phase equations:
\begin{align}
\label{StaEq}
\left\{
\begin{array}{l}
m_u-M_u=G h_u +\frac{\kappa}{16}h_d h_s\\
m_d-M_d=G h_d +\frac{\kappa}{16}h_u h_s\\
m_s-M_s=G h_s +\frac{\kappa}{16}h_u h_d
\end{array}	
\right.,
\end{align}
where $m_i$ stand for the current masses of the quarks, and $M_i$ are the  constituent masses. As mentioned before, we take $m_u=m_d=0$, as in that case the single-particle spectrum from Eq.~(\ref{ESpectrum}) is the exact solution of the Dirac quark Hamiltonian in the light quark sector. The thermodynamical potential respects the usual small-$q$ expansion \cite{Tatsumi:2004dx}
\begin{align}
\Omega_{vac}=\left.\Omega_{vac}\right|_{q=0}+\frac{1}{2}f^2_\pi q^2+\mathcal{O}(q^4),
\end{align}
where $f_\pi$ refers to the pion weak decay constant
($f^2_\pi=M_l^2\frac{N_c J_1(M_l)}{4\pi^2}$, where $J_{1}(M)=\mathrm{ln}\left(1+\frac{\Lambda^2}{M^2}\right)-\frac{\Lambda^2}{\Lambda^2+ M^2}$)
 %\footnote{$f^2_\pi=M_l^2\frac{N_c I_1}{4\pi^2}$}
. 

\section{Results}

\subsection{NJL case}

First, let us consider the ${\rm SU}_2$ NJL model in the chiral limit ($m_u=m_d=0$). In the
usual NJL scenario, when $\tau=\frac{N_c G \Lambda^2}{2\pi^2}>1$, dynamical breaking
of the chiral symmetry is induced and the quarks acquire a finite dynamical mass in
the vacuum \cite{Osipov:2006ns}. Chiral symmetry is restored at a critical value for the chemical
potential (in this paper we consider the $T=0$ case). Let us restrict ourselves to
the family of parameters which result in a fixed value for the vacuum dynamical
mass. For the present results we take
\begin{eqnarray}
M^{vac}=330~\mathrm{MeV},
\end{eqnarray}
which is in the ballpark 
leading to proper meson phenomenology. Together with a choice for
$\tau$ it then determines the values of $G$ and $\Lambda$, the model parameters. 

We have verified that for $\tau>1.23$ a first order transition occurs at a certain critical chemical
potential between the solution with finite mass and the trivial one ($M=0$),
whereas below this critical value a second order transition takes place.

The consideration of the 
%single-particle (quasi-particle) 
energy spectrum given
by Eq.~(\ref{ESpectrum}), resulting in a modification of the thermodynamical potential~(\ref{Omega}),
introduces new rich scenarios (when $q=0$ we recover the usual model). The values of
$h_l$ (the light quark chiral condensate) and the wave vector $q$ are determined by minimizing
the thermodynamical potential. For high enough values of the chemical potential, the global
minimum corresponds to a solution with finite $q$. Asymptotically, this solution,
which corresponds to $\lim_{\mu\rightarrow\infty}\left\{h,q\right\}=\{0,2\mu\}$,
becomes degenerate with the trivial one. It is worth pointing out that when the
dynamical mass goes to zero, the thermodynamical potential becomes independent of
the value of $q$, $\frac{\partial}{\partial q}\Omega\left(M=0,q\right)=0$, as no condensates are present 
in this case.

\begin{figure*}
\includegraphics[width=0.75\textwidth]{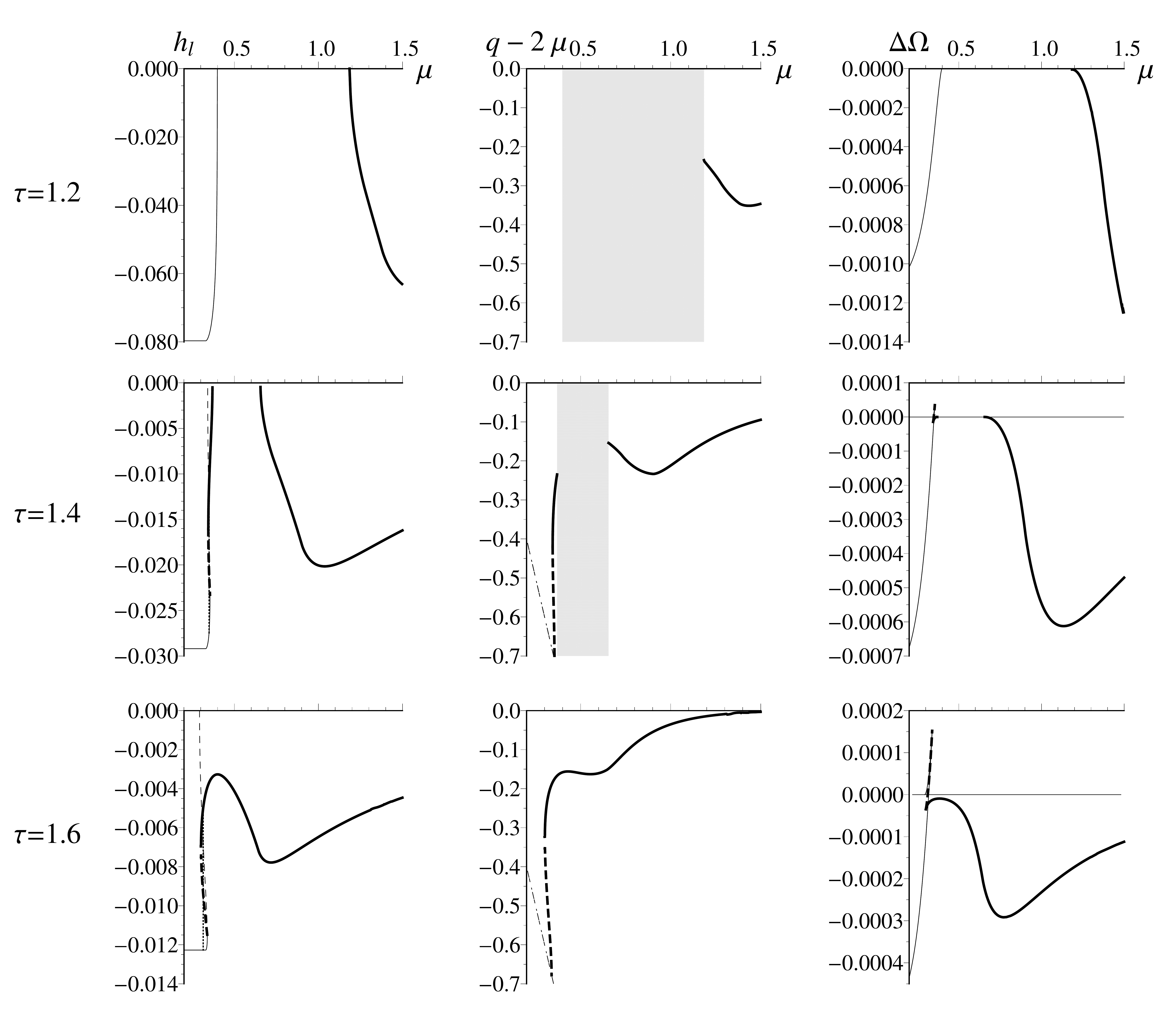}
\caption{In each row, from left to right, we show the chemical potential dependence of
the solutions for the light quarks chiral condensate, the wavelength of the
modulation of the condensates (here we plot the difference $q-2\mu$ to emphasize
that $q$ goes asymptotically to $2\mu$), and the thermodynamical potential
difference with respect to the trivial solution for the given values of the model parameter $\tau$.
Thicker lines correspond to finite-$q$ solutions. In all plots of the paper 
we use the units $[h]=\mathrm{GeV}^3$, $[\mu]=\mathrm{GeV}$, and
$[\Omega]=\mathrm{GeV}^4$. The gray area corresponds to the interval where the
value of $q$ is undetermined, as $M=0$ (see the main text for details).}
\label{grafpainelNJLSU2}
\end{figure*}

In the first row of Fig. \ref{grafpainelNJLSU2} we can see an example of the
%crossover case 
second-order phase transition ($\tau=1.2$). Besides the usual finite condensate solution, which
merges with the trivial solution at a critical value of the chemical potential
(in this case $\mu_c=399~\mathrm{MeV}$) another solution corresponding to a
global minimum appears at a much higher chemical potential
($\mu_c=1.187~\mathrm{GeV}$). Asymptotically it becomes degenerate with the
trivial solution.

\begin{figure*}
\center
\includegraphics[width=0.75\textwidth]{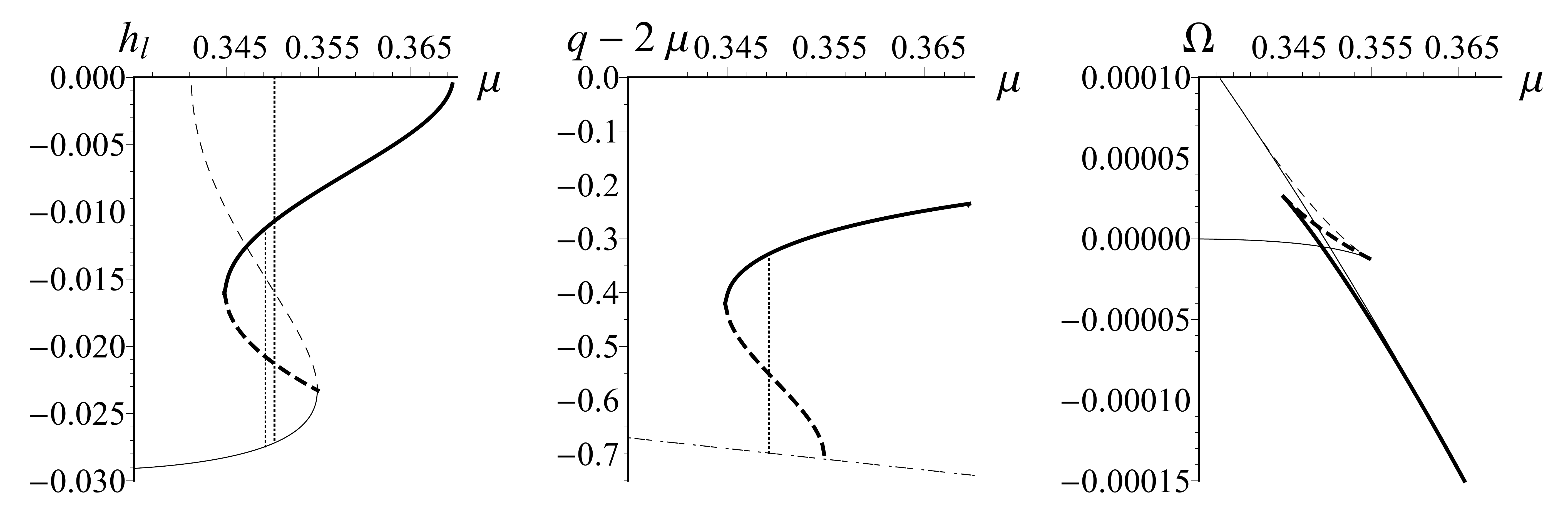}
\caption{
A zoom for chemical potential dependence of $h_l$, $q-2\mu$, and $\Omega$ around
the first order transition point for the $\tau=1.4$ case (the same as in the second
row of Fig.~\ref{grafpainelNJLSU2}). In the leftmost panel we can see the merging
of three lines, which corresponds to the annihilation of a minimum, a maximum, and
a saddle point of $\Omega(h_l,q)$. The two vertical dotted lines mark the jump
(the rightmost is the usual jump when we consider only $q=0$ and the leftmost
the jump to the finite $q$ solution). As before, we use thicker lines to denote
the finite $q$ solutions. The diagonal straight dot-dashed line represents
$q=0$.
}
\label{grafzoomsNJLSU2tau14}
\end{figure*}

For values of $1.23<\tau <1.53$ we get a first order transition at a critical
chemical potential between the solution with a finite mass and vanishing $q$ and a
finite mass solution with finite $q$ (see second row in Fig.~%
\ref{grafpainelNJLSU2} and Fig.~\ref{grafzoomsNJLSU2tau14}). This transition
occurs at a chemical potential slightly below that of the 
%traditional 
usual first
order transition to the trivial solution. This branch disappears for a
higher value of the chemical potential and as a result there is a chemical
potential window before the appearance of the solution similar to the one
described above (with $q\rightarrow 2\mu$) where the chiral symmetry is restored.

For even higher values, $\tau>1.53$, these solution branches meet and for any
chemical potential above the transition the global minimum  corresponds to a
solution with a finite mass and $q$ (see the third row in Fig.~%
\ref{grafpainelNJLSU2}). Thus chiral symmetry is restored only asymptotically.

\begin{figure*}
\center
\subfigure[]{\label{grafPotQuimCritVsTauNJLCW}\includegraphics[width=0.24\textwidth]{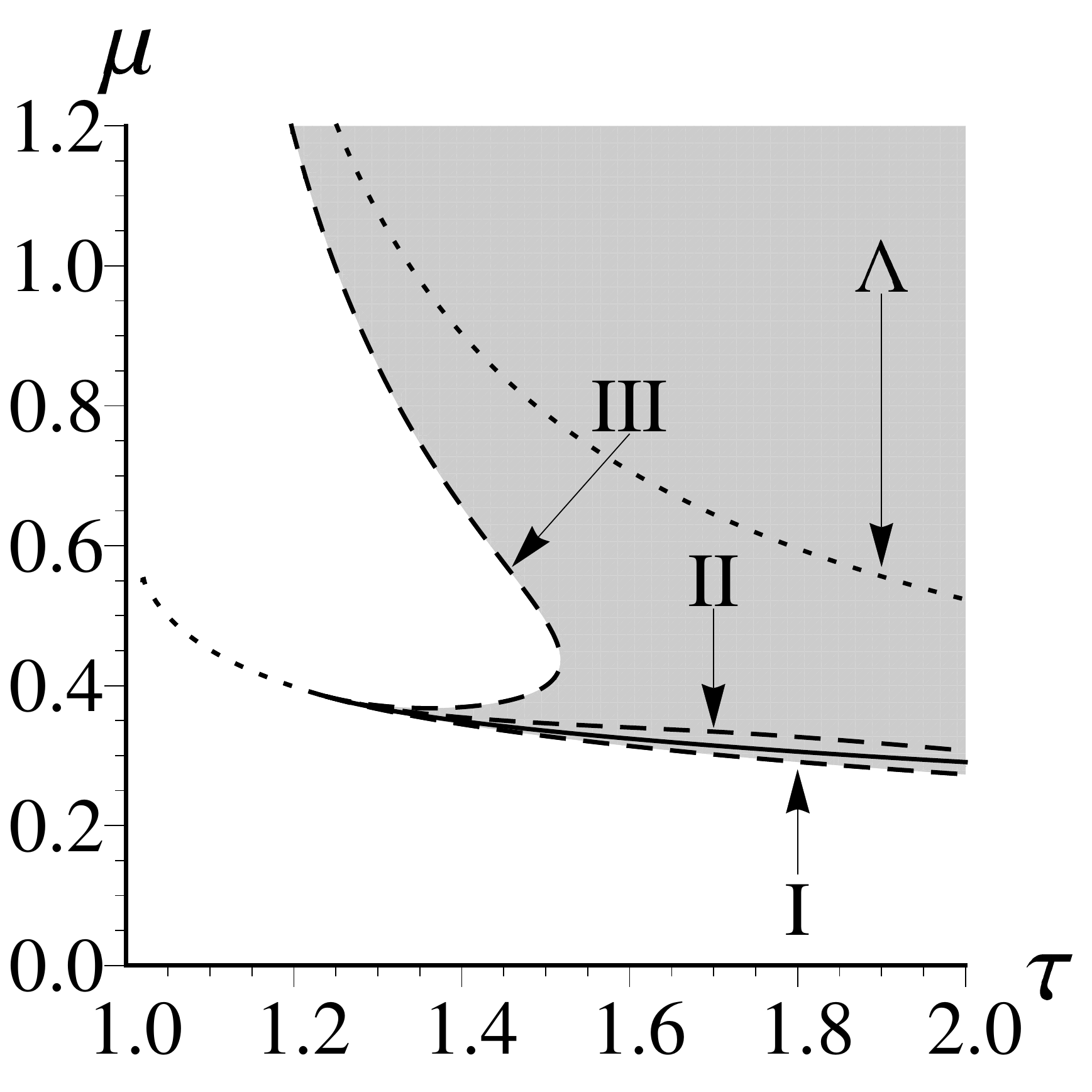}}
\subfigure[]{\label{grafAuxPotQuimSolNJLCWSU2}\includegraphics[width=0.25\textwidth]{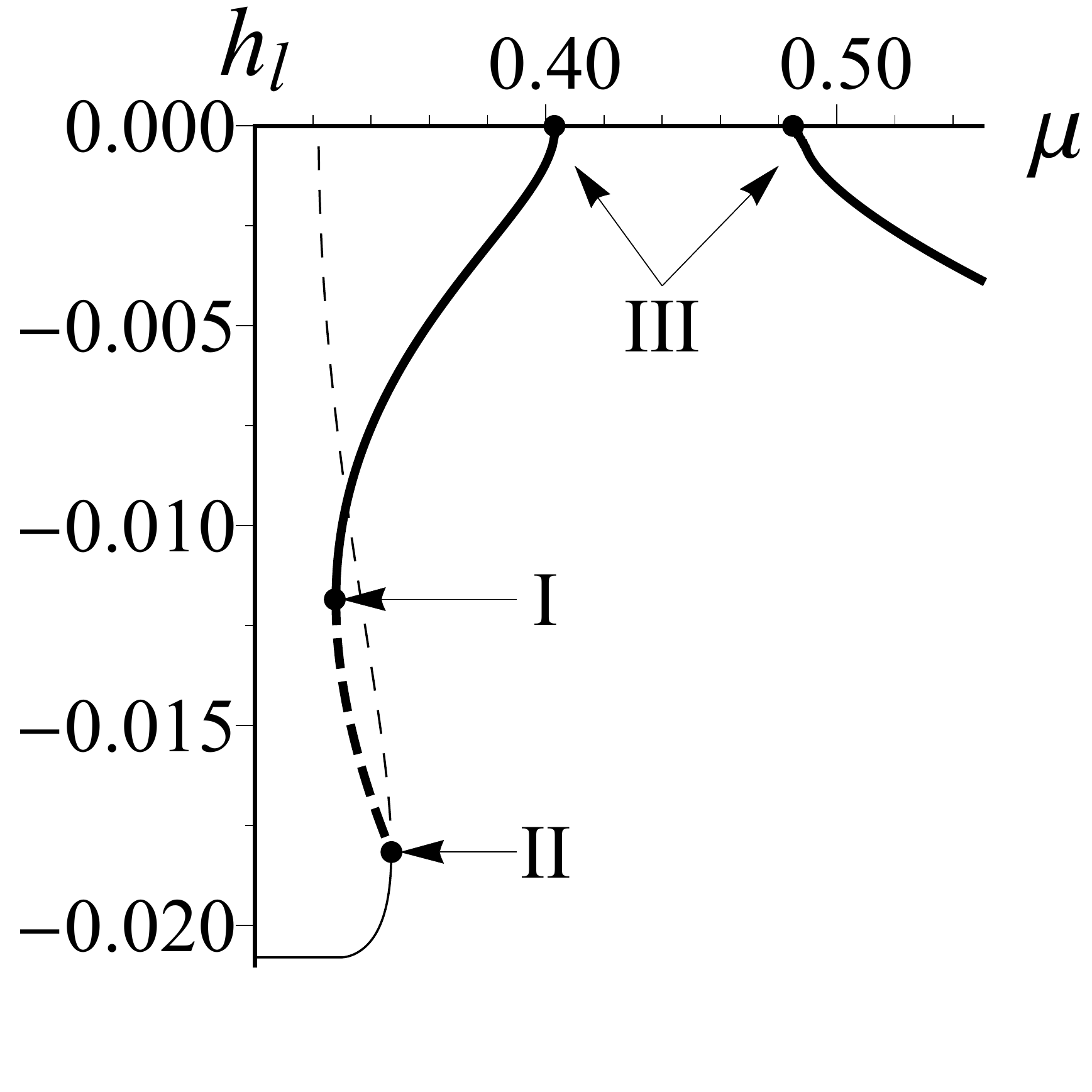}}%\\
\subfigure[]{\label{grafqCritVsTauNJLCW}\includegraphics[width=0.24\textwidth]{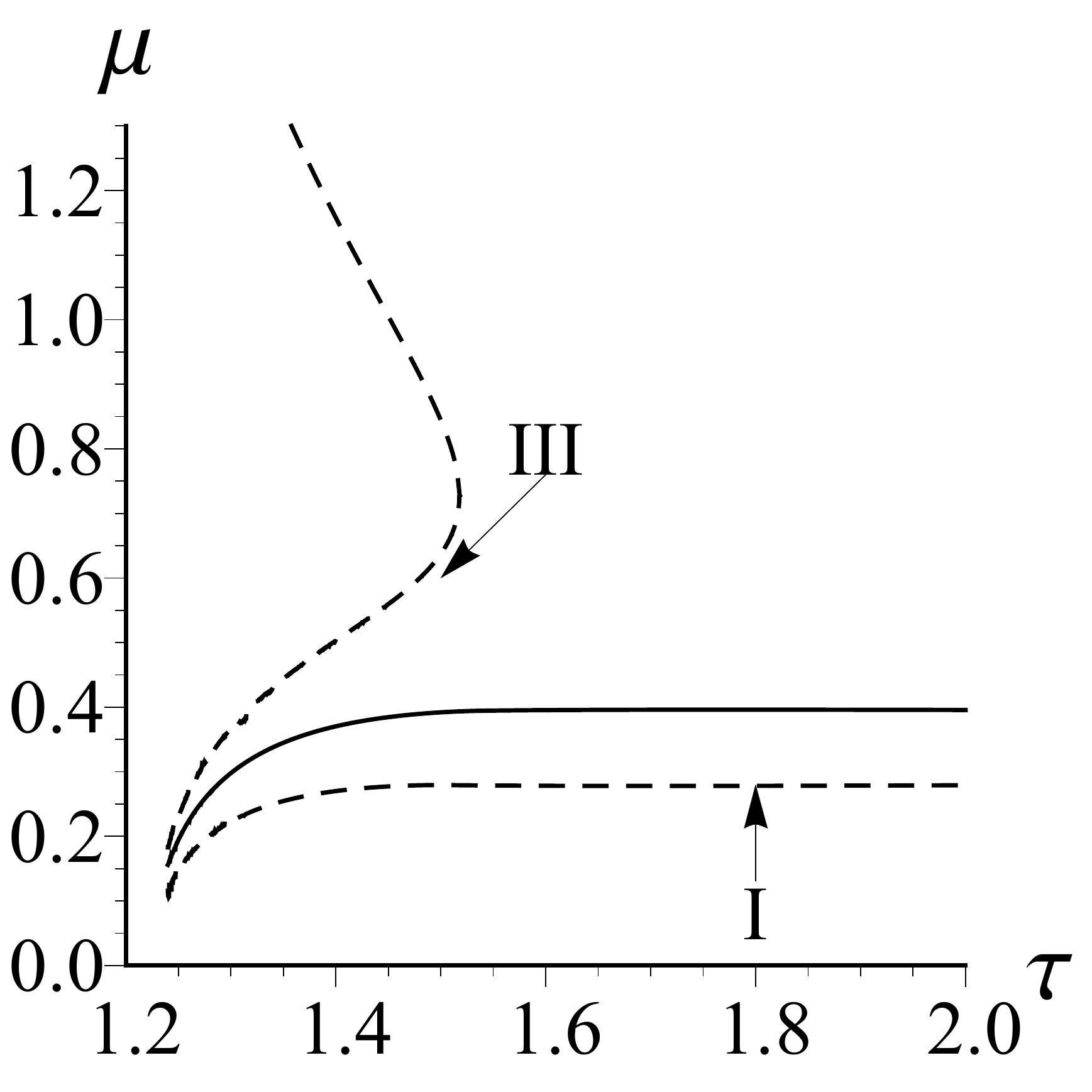}}
\subfigure[]{\label{grafhCritVsTauNJLCW}\includegraphics[width=0.24\textwidth]{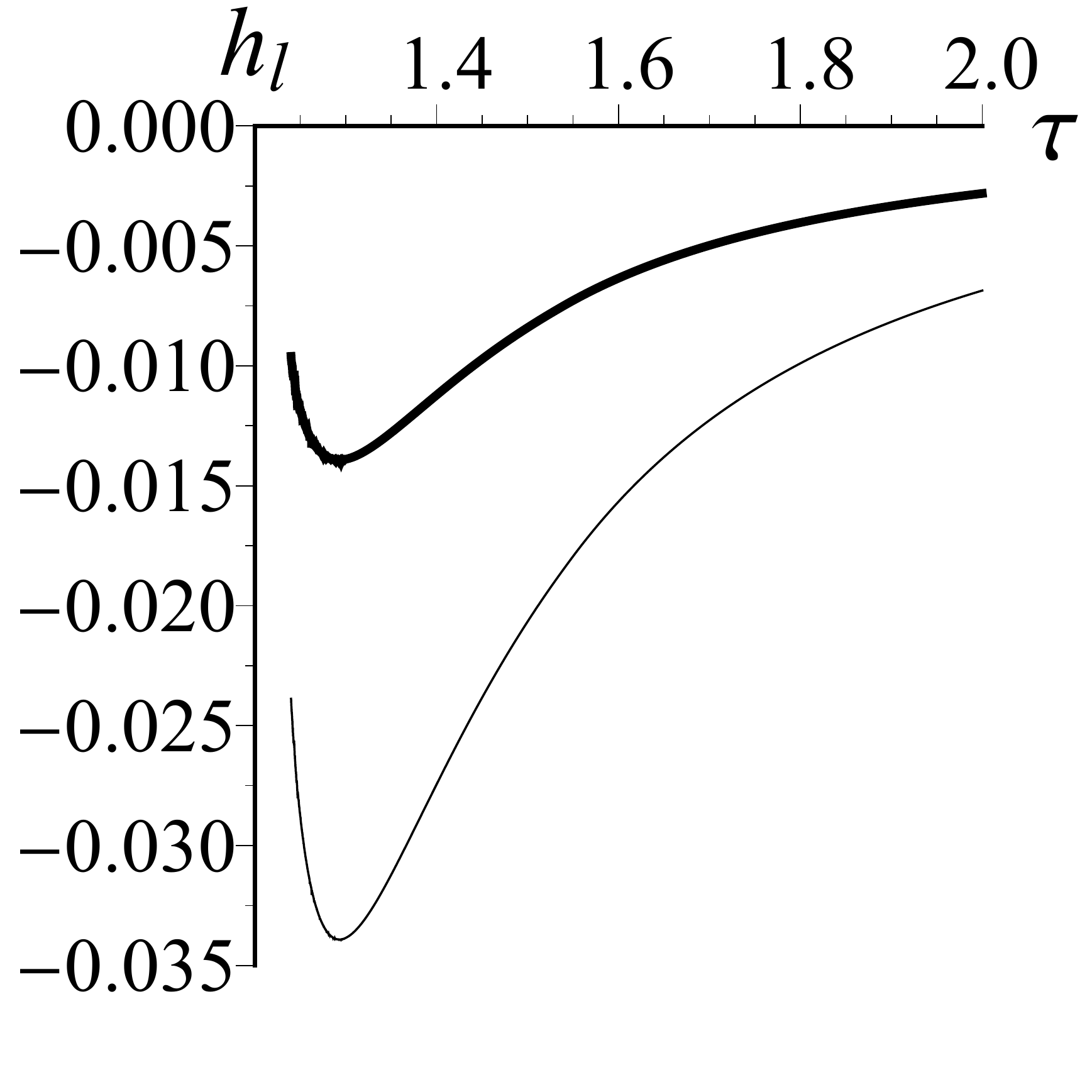}}
\caption{In \ref{grafPotQuimCritVsTauNJLCW} we present the critical chemical potentials as
a function of $\tau$ in the NJL model. For $\tau<1.23$ there are two
transitions: a crossover between the dynamically broken phase and the trivial
phase (marked the dotted line), and, at a higher chemical potential, a transition to
a phase with finite $q$ (type III, dashed line). For $\tau>1.23$ a transition to
a phase with finite $q$  replaces the crossover transition. It is represented by the full
line. For $1.23<\tau<1.53$ this finite-$q$ solution disappears at a critical
chemical potential (marked by the lower part of the type III dashed line) but at
an even higher chemical potential the preferred solution becomes again the one with
finite $q$ (upper part of the type III dashed line). The shaded area marks
the delimitation of the finite $q$ phase. The top-most dotted line corresponds to
the choice of $\Lambda$. In \ref{grafAuxPotQuimSolNJLCWSU2} we plot the
condensate solutions as functions of the chemical potential for the case of
$\tau=1.5$ to illustrate what is meant by types I, II and III (type I finite q and h;
type II vanishing q and finite h; type III vanishing h and finite q).  In
\ref{grafqCritVsTauNJLCW} we show the value of $q$ for the finite $q$ solutions involved
in the critical points (the line types are the same as in Fig.
\ref{grafPotQuimCritVsTauNJLCW}). In \ref{grafhCritVsTauNJLCW} we show the condensate for
the solutions with finite and vanishing $q$ (thicker for finite $q$) at the
first order transition (the one which is indicated in
\ref{grafPotQuimCritVsTauNJLCW} by the full line).}
\label{CritVsTau}
\end{figure*}

The critical values of the chemical potentials described above are shown as a function of
$\tau$ in Fig.~\ref{CritVsTau}.

\subsection{NJLH case}
Now let us consider the extension of the model to include the strange quark and
the effect of the 't Hooft determinant. As before, we choose the model parameters
in such a way as to obtain $M^{vac}_l=330~\mathrm{MeV}$. Furthermore, we chose to
fix the value of $\tau$ at $1.4$. In order to better understand and
differentiate the effects of flavor mixing and of the inclusion of a finite
current quark mass, we consider both the case of a realistic value for the strange
current mass and the chiral limit. It should be noted that in the case with
the finite current mass for the strange quark the value of its dynamical mass in the
vaccum depends on the coupling constant $\kappa$ ($M^{vac}_s=586~\mathrm{MeV}$ for
$\kappa=0$ and $M^{vac}_s=542~\mathrm{MeV}$ for
$\kappa=-2000~\mathrm{GeV}^{-5}$).

\subsubsection{Finite strange quark current mass}

Without the OZI-violating 't Hooft determinant term  ($\kappa=0$), the light and
strange sectors are decoupled. The solutions obtained for $h_l$ are, as such,
the same as in the second row of Fig.~\ref{grafpainelNJLSU2}. An additional
first order transition (at a critical chemical potential close to $M^{vac}_s$)
appears, resulting in a jump in the value of $h_s$ (in this section the value
considered for the current mass of the quark strange is $m_s=186~\mathrm{MeV}$).

\begin{figure*}
\includegraphics[width=0.75\textwidth]{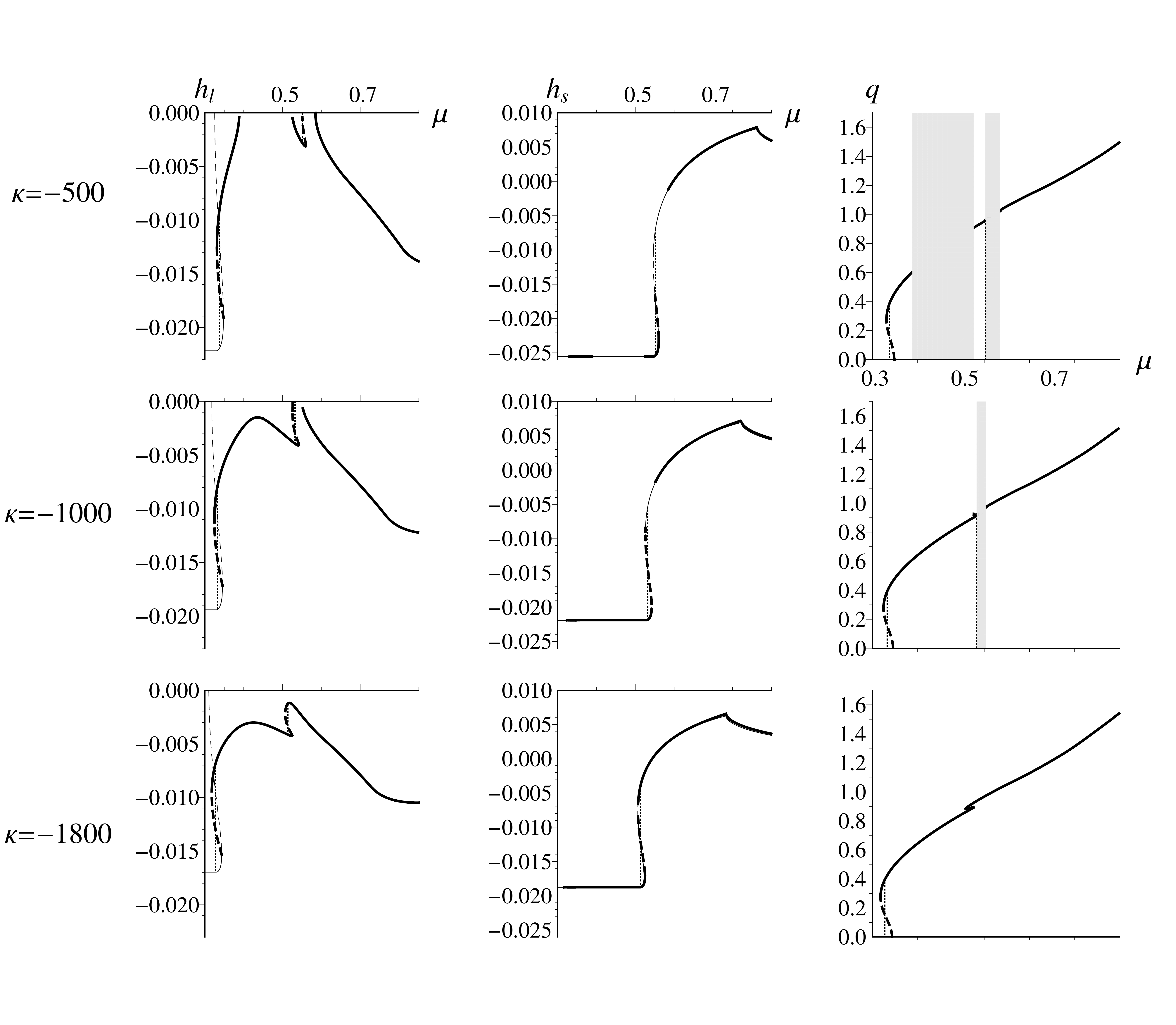}
\caption{In each row, from left to right, we present the chemical potential dependence of
the solutions for the light and strange quark chiral condensate and for the
wavelength of the modulation of the condensates, for coupling strengths of the
'tHooft determinant ($\left[\kappa\right]=\mathrm{GeV}^5$). Thicker lines
correspond to the finite-$q$ solutions. The gray area corresponds to the interval
where the value of $q$ is undetermined, since $M=0$.}
\label{grafpainelNJLHSU3}
\end{figure*}

Turning on flavor mixing couples the gap equations for the light and
strange sectors. Depending on the coupling strength of the 't Hooft determinant,
we can get several different scenarios (see Fig.~\ref{grafpainelNJLHSU3}).

\begin{figure*}
\subfigure[]{\label{grafSolhlNJLHtau14K500zoomI}\includegraphics[width=0.24\textwidth]{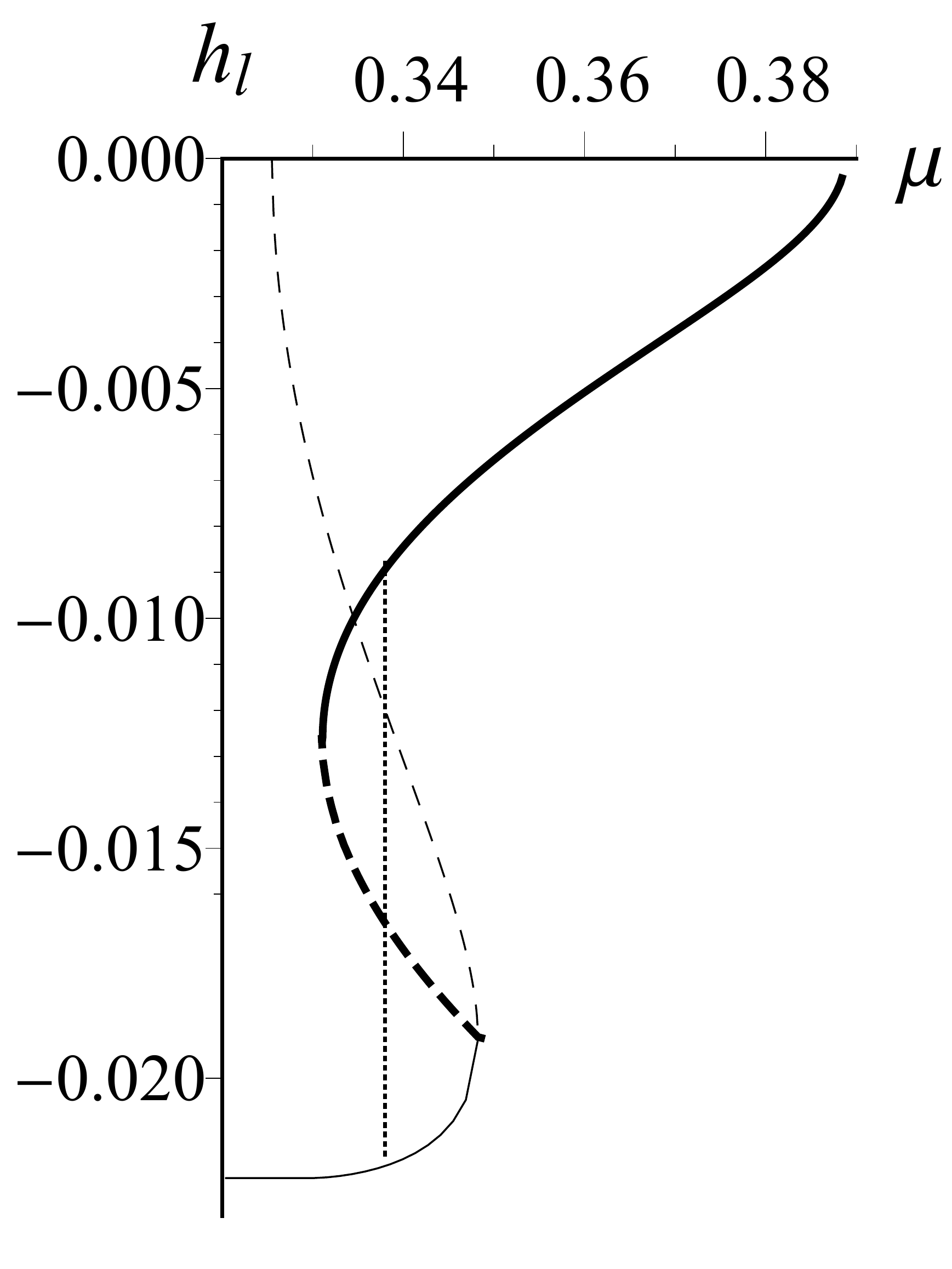}}
\subfigure[]{\label{grafSolhsNJLHtau14K500zoomI}\includegraphics[width=0.24\textwidth]{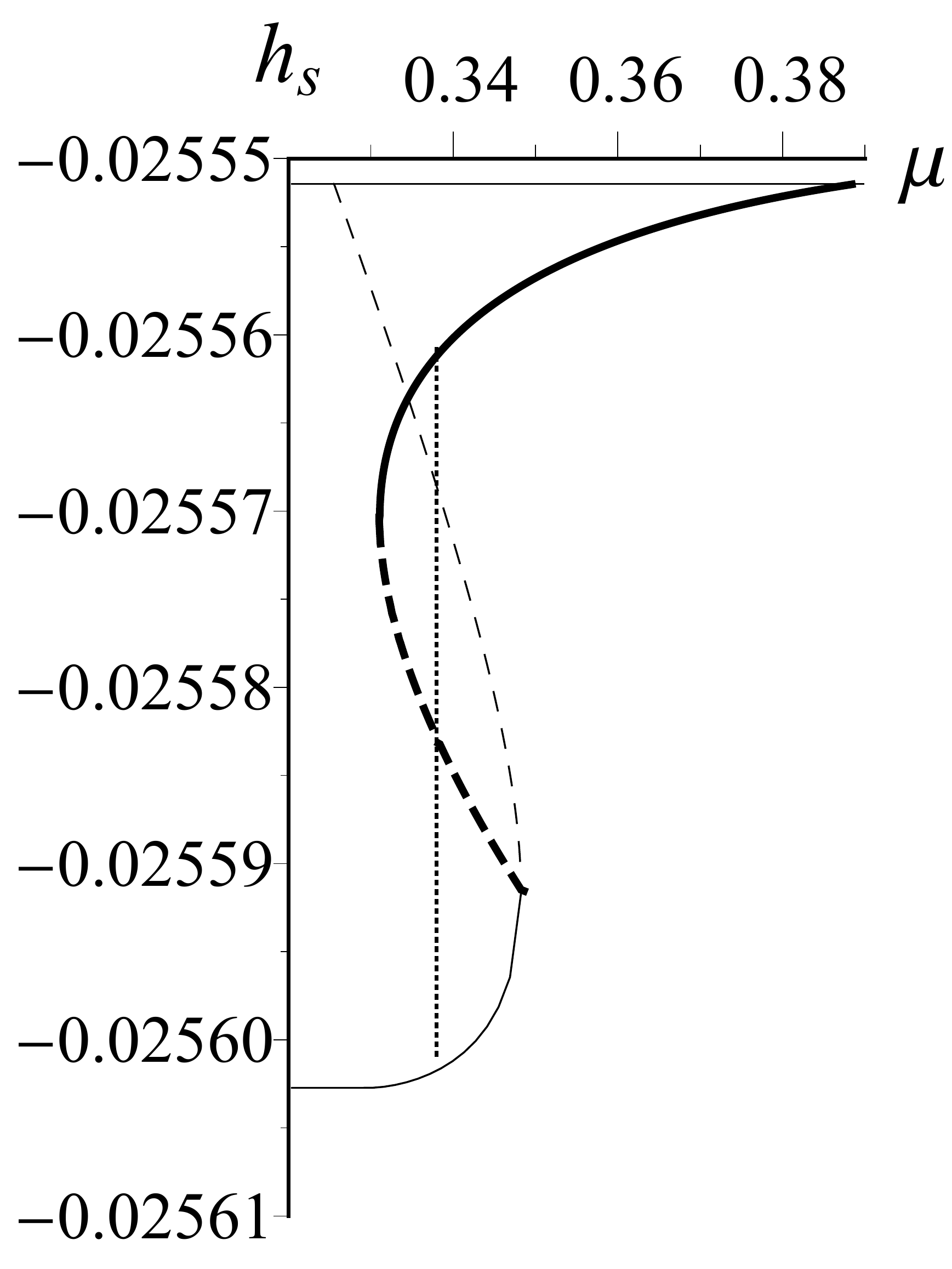}}
\subfigure[]{\label{grafSolhlNJLHtau14K500zoomII}\includegraphics[width=0.24\textwidth]{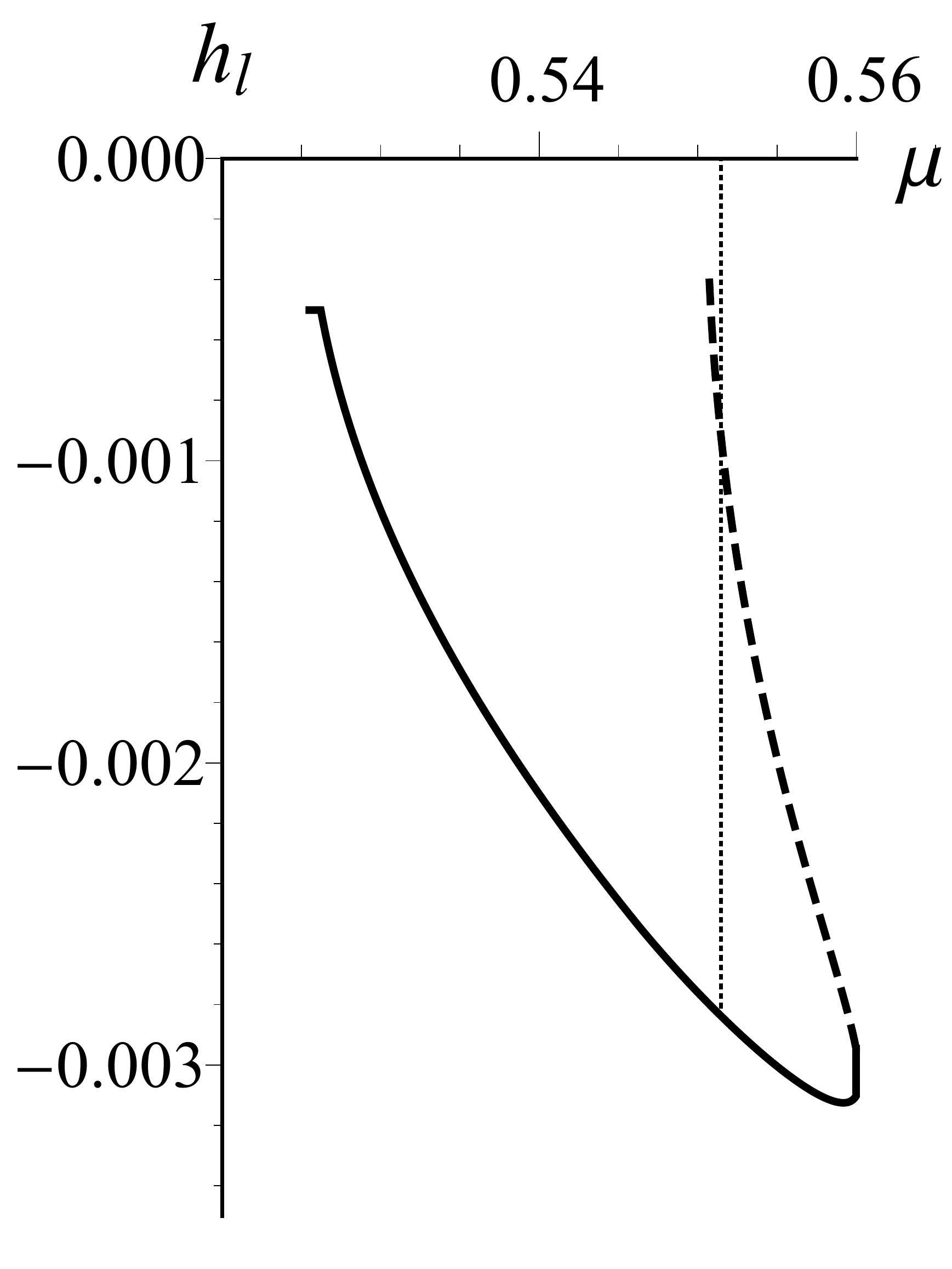}}
\subfigure[]{\label{grafSolhsNJLHtau14K500zoomII}\includegraphics[width=0.24\textwidth]{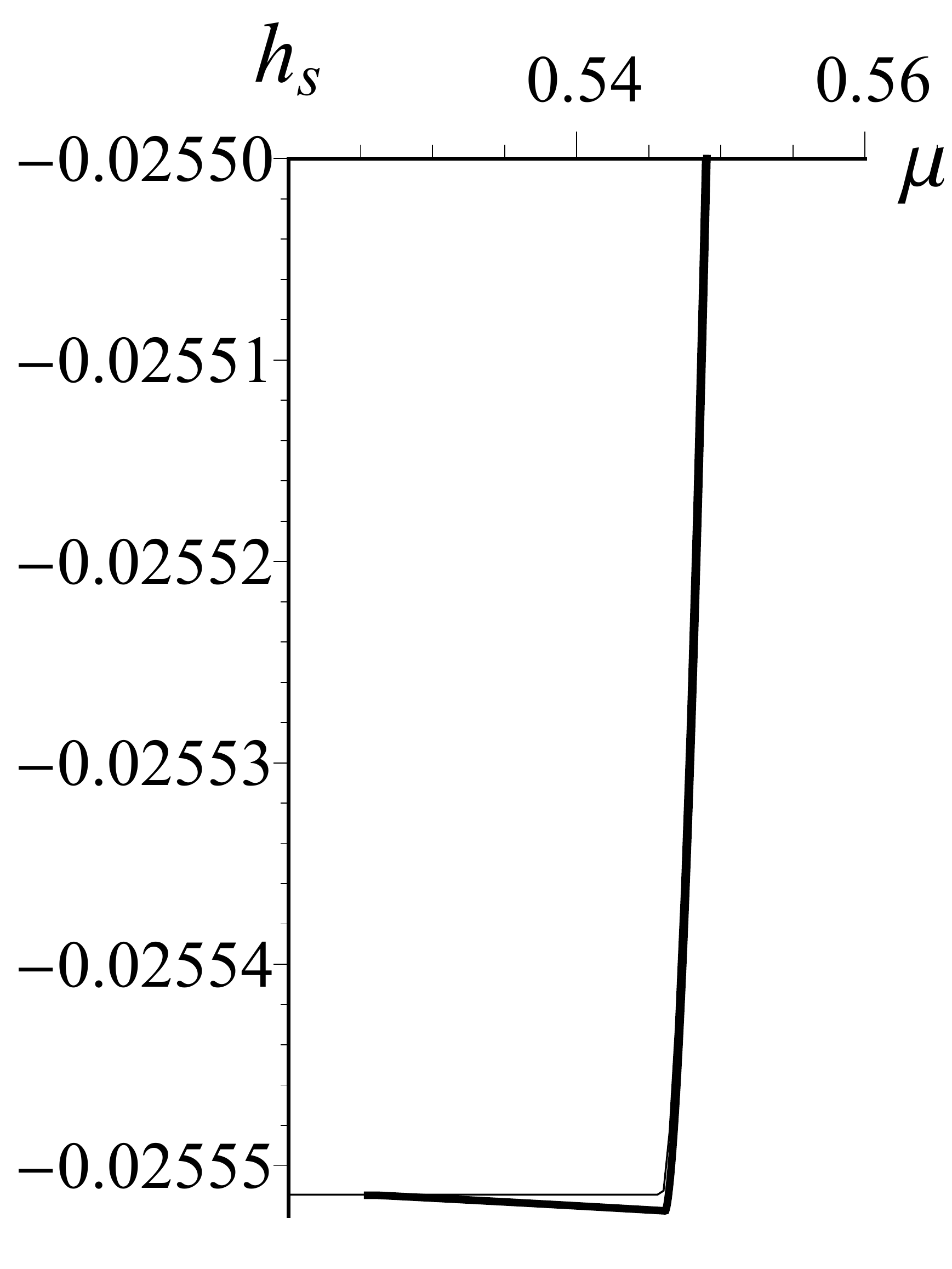}}
\caption{
A zoom for the chemical dependence of the chiral condensate solutions (thicker
lines refer to finite-$q$ solutions) near the transitions (marked by the
vertical dotted lines), presented for the case with
$\kappa=-500~\mathrm{GeV}^{-5}$.}
\label{SolhlsNJLHtau14K500zoom}
\end{figure*}

For $-\kappa>290~\mathrm{GeV}^{-5}$ a new solution branch appears with a shark-fin 
shape for the light condensate in the vicinity of the chemical
potential corresponding to the vacuum dynamical mass of the strange quark. In
the first row of Fig. \ref{grafpainelNJLHSU3} we present some results obtained
by considering $\kappa=-500~\mathrm{GeV}^{-5}$, where three separate chemical
potential windows with finite value of $q$ appear. With increasing chemical
potential we go through two first-order transitions and three crossovers. The
latter involve the dissapearance or emergence of the light condensate, as well as going
to or from a finite $q$ solution to indeterminate $q$. The two first order
transitions occur slightly before (for the one occuring near $M_l^{vac}$) and
slightly above (near $M_s^{vac}$), thus excluding the occurrence of the $q=0$
transitions. A zoom of the behavior of the chiral condensates near the
transitions can be seen in Fig.~\ref{SolhlsNJLHtau14K500zoom}.

For a 't Hooft interaction term stronger than $-\kappa>935~\mathrm{GeV}^{-5}$ the
first two chemical potential windows with finite-$q$ solutions merge, resulting in
the disappearance of the corresponding crossover transitions, as can be seen in
the second row of Fig.~\ref{grafpainelNJLHSU3}. 

With $-\kappa>1660~\mathrm{GeV}^{-5}$ the last transition to a vanishing $q$ solution
does not occur and is substituted by a first-order transition between two phases
with finite $q$ (see the third row of Fig.~\ref{grafpainelNJLHSU3}).

\begin{figure*}
\center
\subfigure[]{\label{grafPotQuimCritVsKNJLHCWtau14shade}\includegraphics[width=0.24\textwidth]{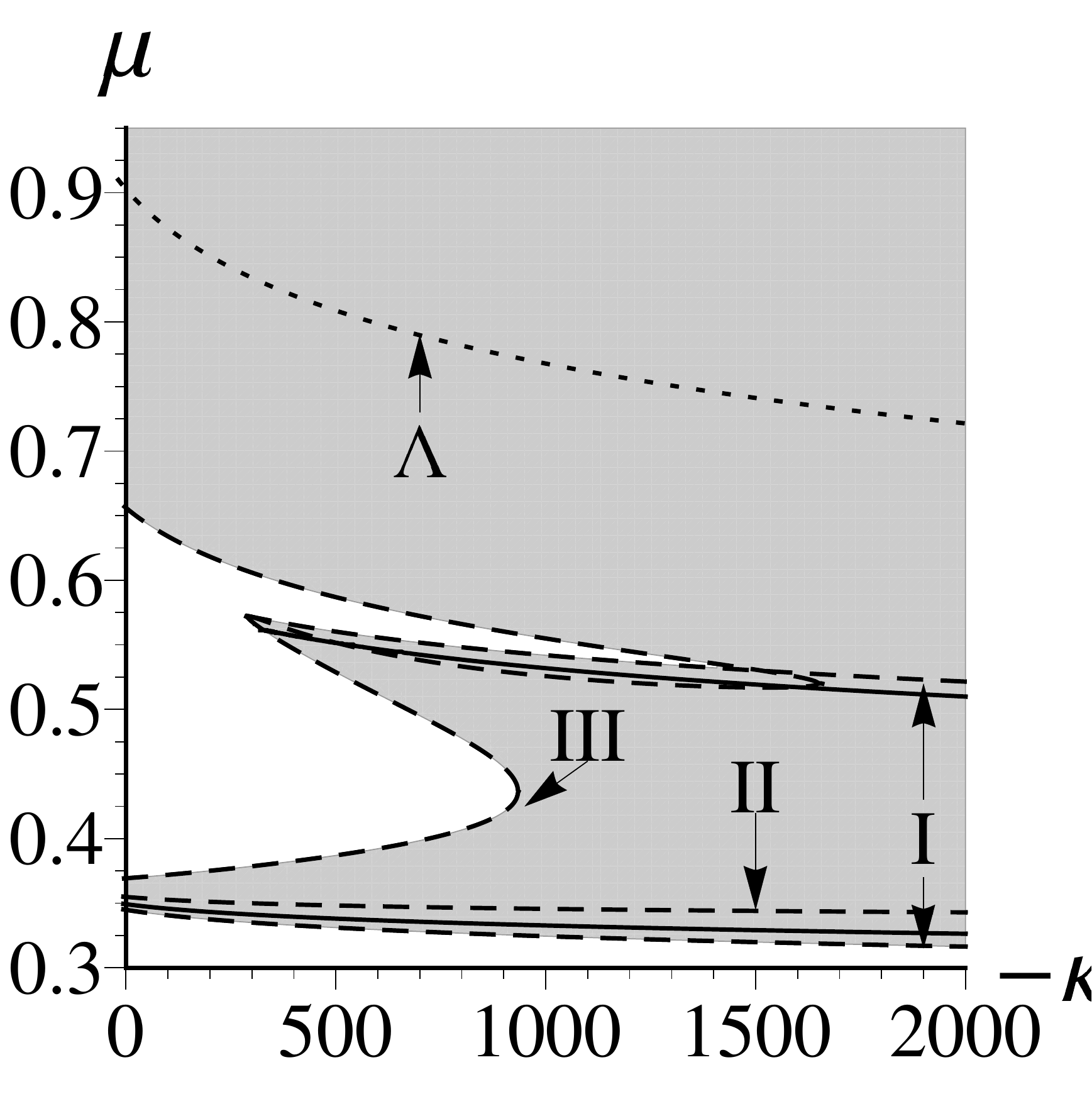}}
\subfigure[]{\label{grafAuxPotQuimSolNJLHCW}\includegraphics[width=0.24\textwidth]{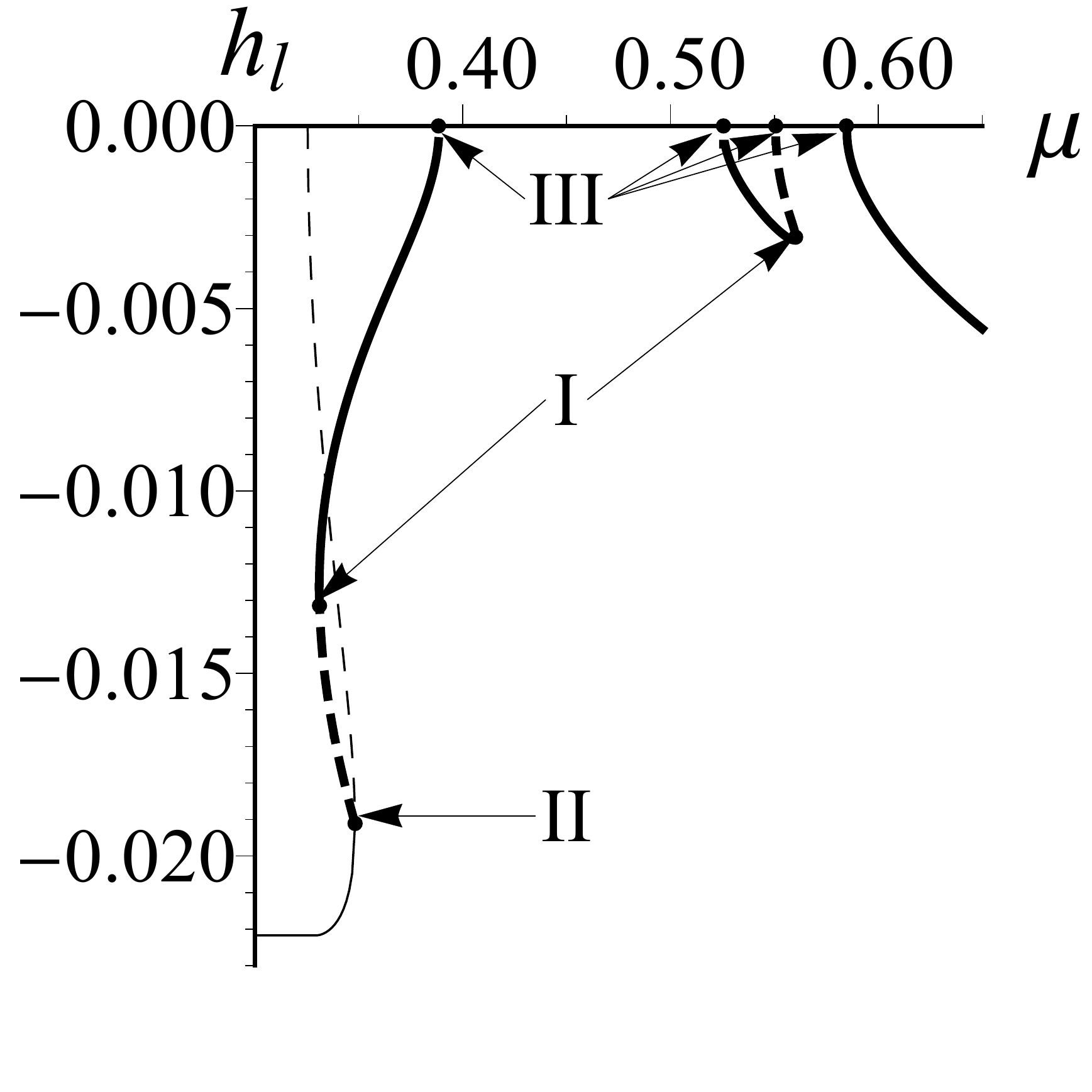}}
\caption{Critical chemical potentials as a function of the intensity of the 't
Hooft determinantal term ($[\kappa]=\mathrm{GeV}^5$). The upper dotted
lines corresponds to the choice of the cutoff. The chemical potential of the first
order transitions are marked by the full black lines. Dashed lines indicate the
borders of the region where the finite-$q$ solutions exist. We distinguish between
three types of critical chemical potentials and an example for this distinction
in the $\kappa=-500~\mathrm{GeV}^{-5}$ case appears in
Fig.~\ref{grafAuxPotQuimSolNJLHCW} (type I with finite $q$ and $h_l$, type II with
finite $h_l$ and vanishing $q$, and type III with finite $q$ but vanishing
$h_l$).}
\label{muCritVsK}
\end{figure*}

The dependence on the 't Hooft coupling strength of the critical chemical potentials
can be seen in Fig.~\ref{muCritVsK}. 

\begin{figure*}
\center
\subfigure[]{\label{grafSolqNJLHtau14K500zoom}\includegraphics[width=0.24\textwidth]{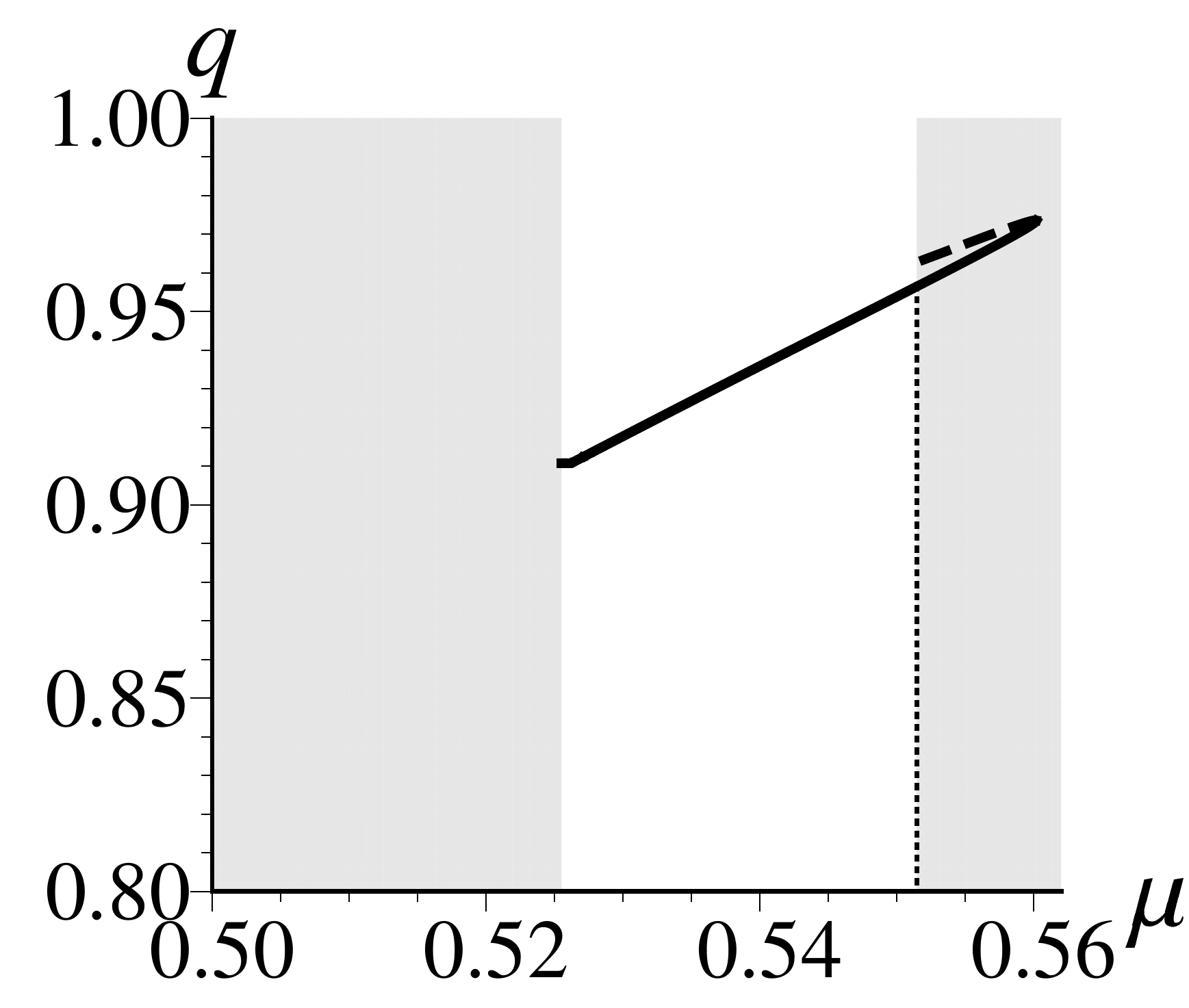}}
\subfigure[]{\label{grafSolqNJLHtau14K1000zoom}\includegraphics[width=0.24\textwidth]{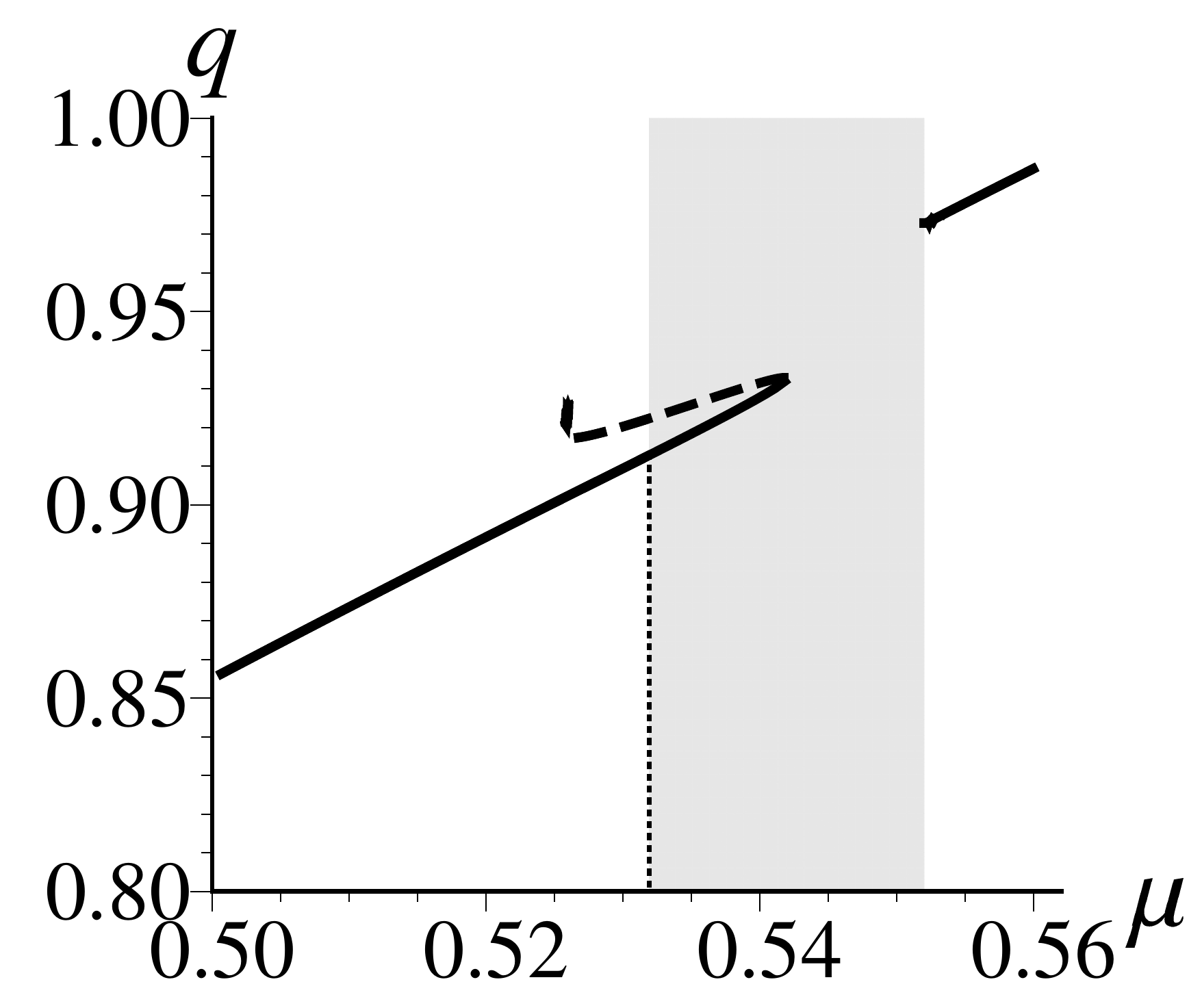}}
\subfigure[]{\label{grafSolqNJLHtau14K1800zoom}\includegraphics[width=0.24\textwidth]{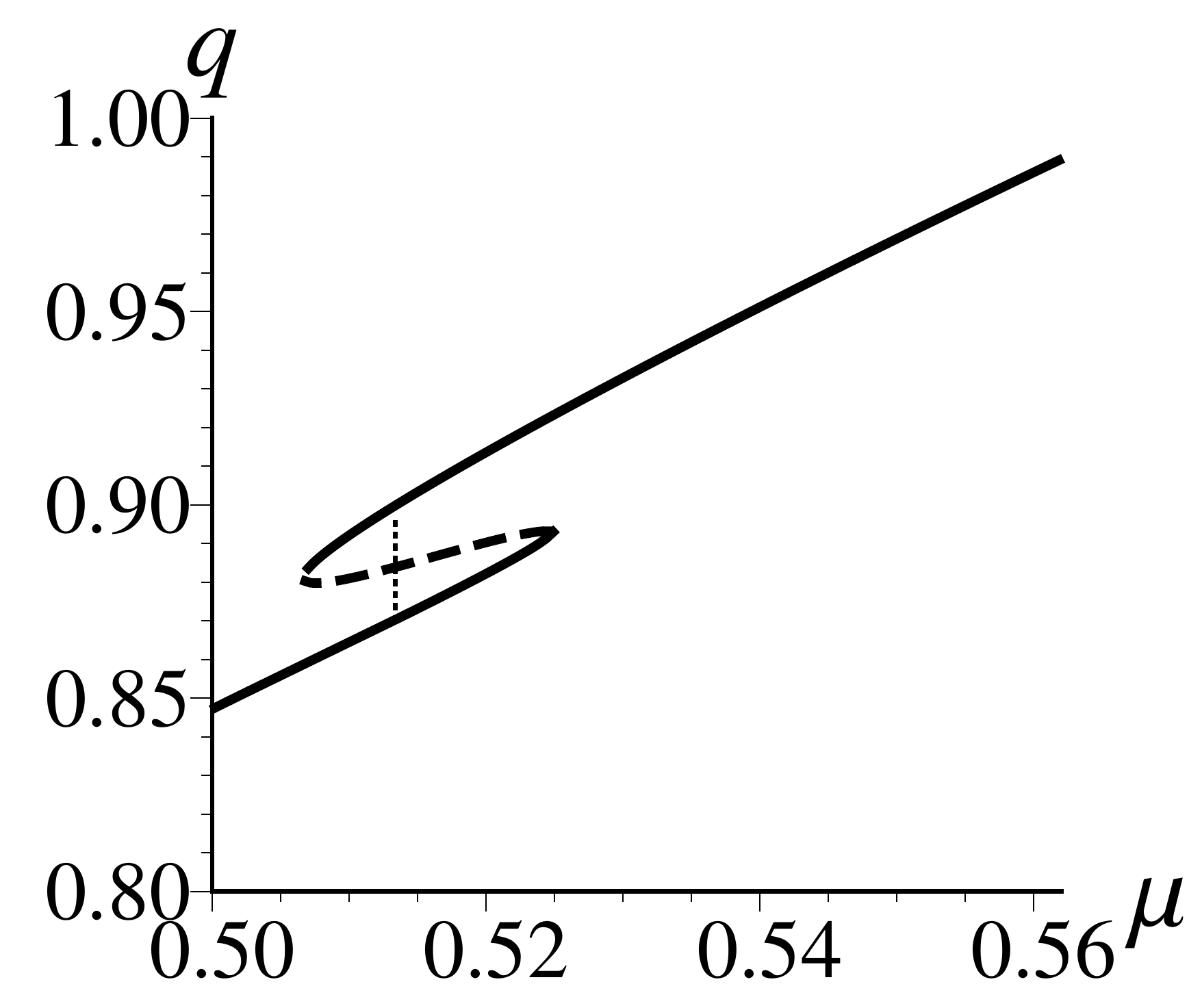}}
\caption{Zoom of the solutions for $q$ in the chemical potential window close to
$M^{vac}_s$ for $\kappa=-500,~-1000,~-1800~\mathrm{GeV}^5$ (from left to right),
showing the merging of the solution branches for strong enough flavor mixing.}
\label{grafSolNJLHtau14K1000}
\end{figure*}

\subsubsection{${\rm SU}_3$ Chiral limit}
\begin{figure*}
\begin{center}
\subfigure[]{\label{grafSolhlNJLHtau14K100chilimbw}\includegraphics[width=0.24\textwidth]{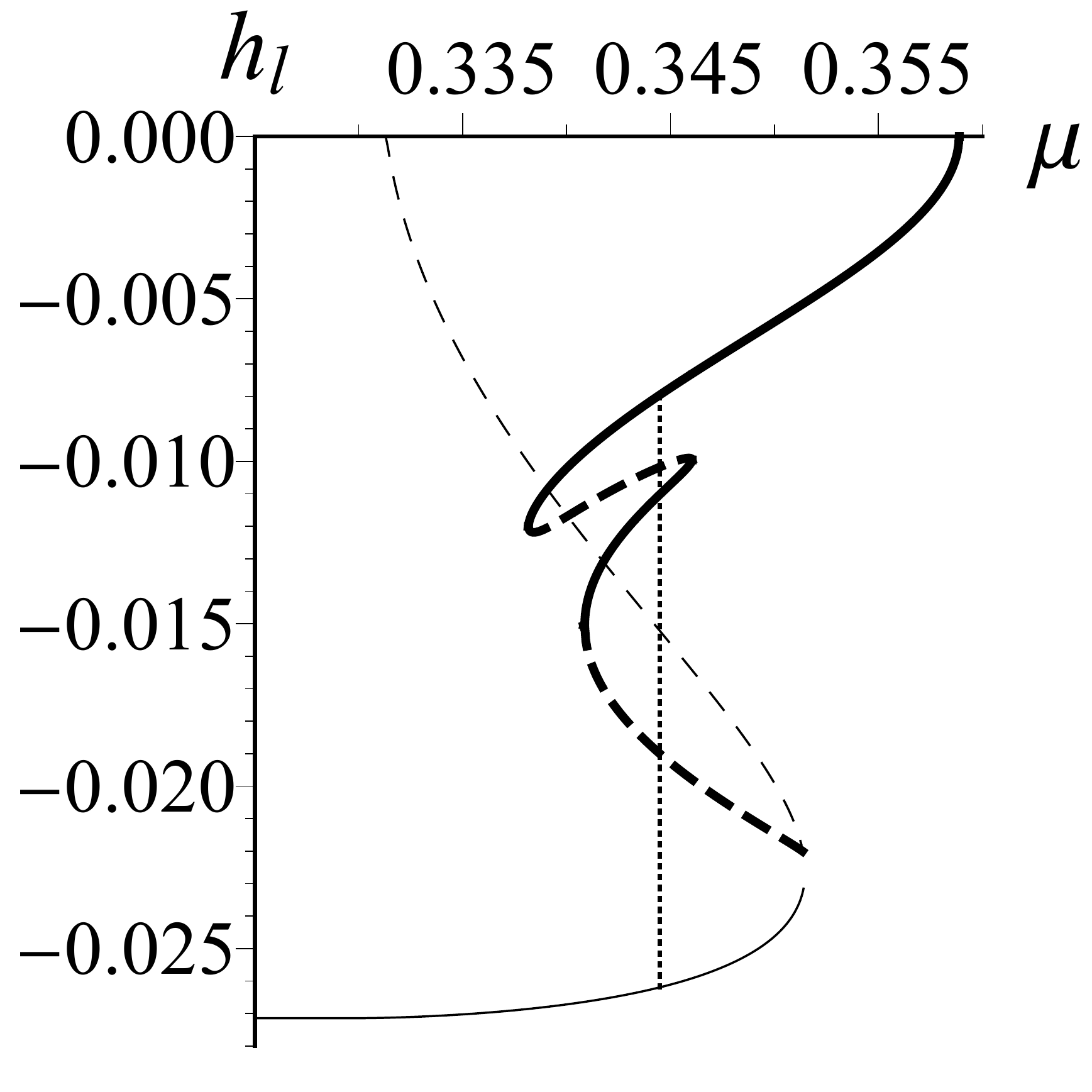}}
\subfigure[]{\label{grafSolhsNJLHtau14K100chilimbw}\includegraphics[width=0.24\textwidth]{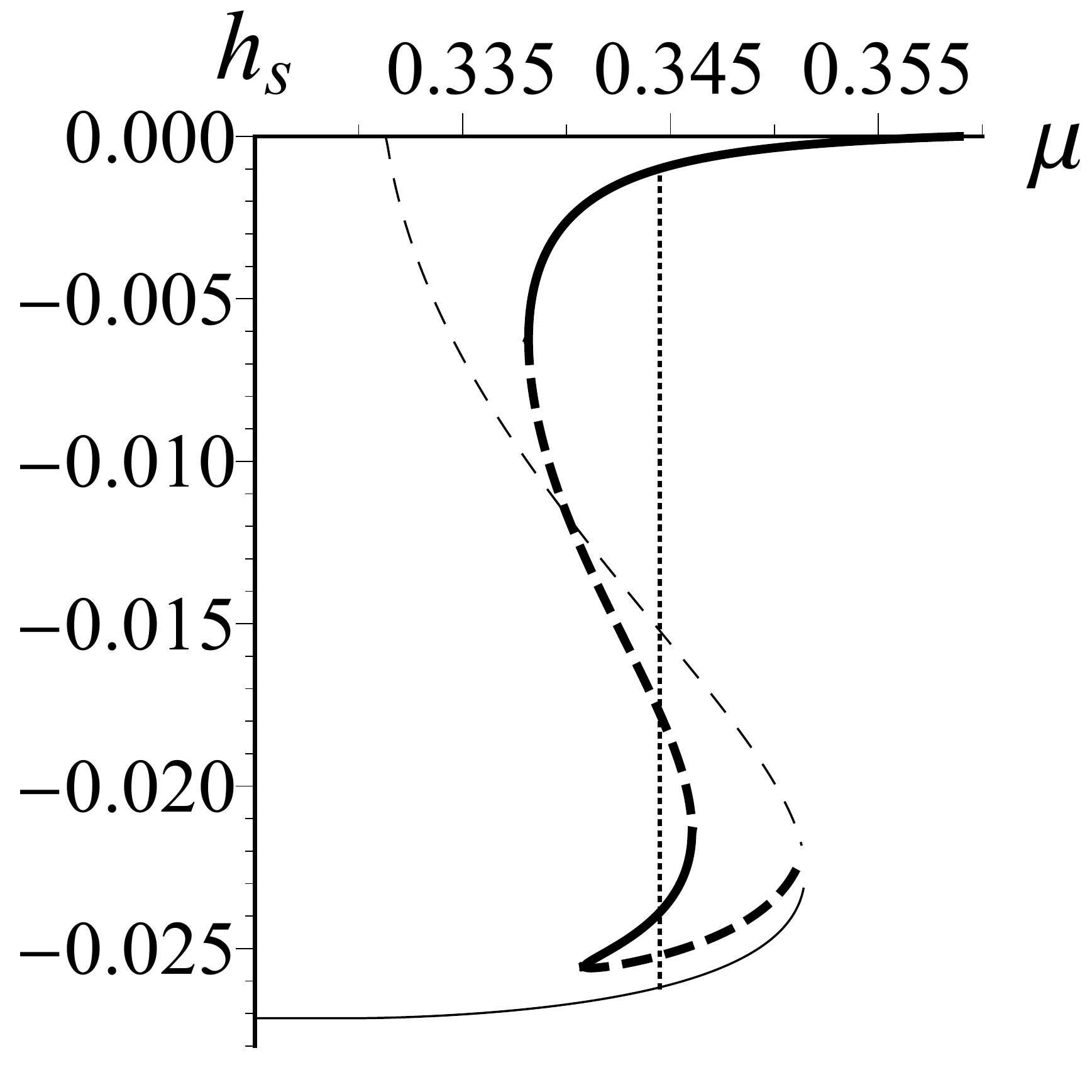}}
\subfigure[]{\label{grafSolqNJLHtau14K100chilimbw}\includegraphics[width=0.24\textwidth]{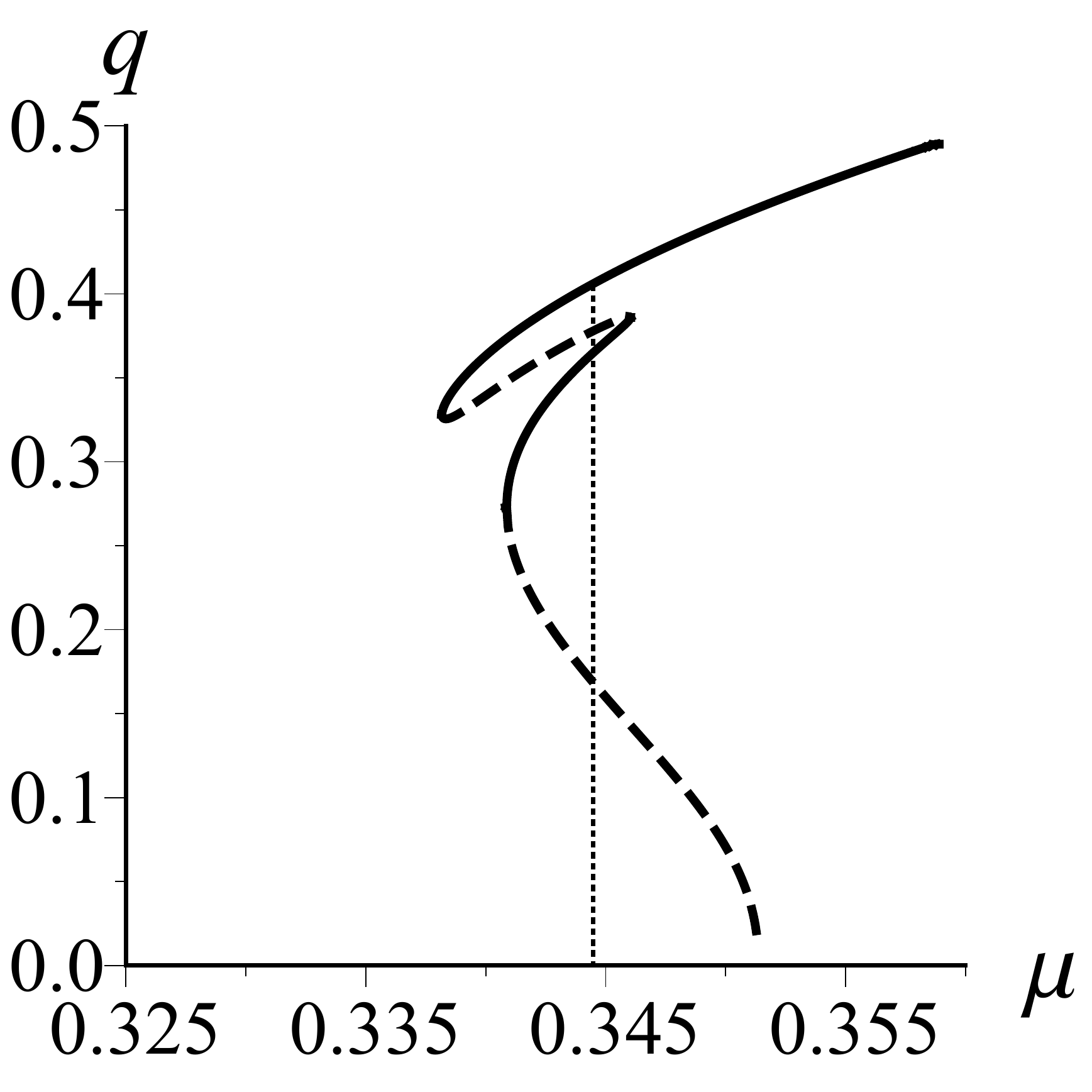}}
\caption{From left to right, we show the chemical potential dependence of the solutions for the light
chiral condensate, the strange chiral condensates (thicker lines with finite
$q$), and of the wavelength of the modulation $q$. The case corresponds to the chiral limit,
$\tau=1.4$, and $\kappa=-100~\mathrm{GeV}^{-5}$. The vertical dotted line marks the first-order transition.}
\label{SolNJLHtau14K100chilim}
\end{center}
\end{figure*}

\begin{figure*}
\begin{center}
\subfigure[]{\label{grafPotQuimCritVsKNJLHCWtau14chilim}\includegraphics[width=0.24\textwidth]{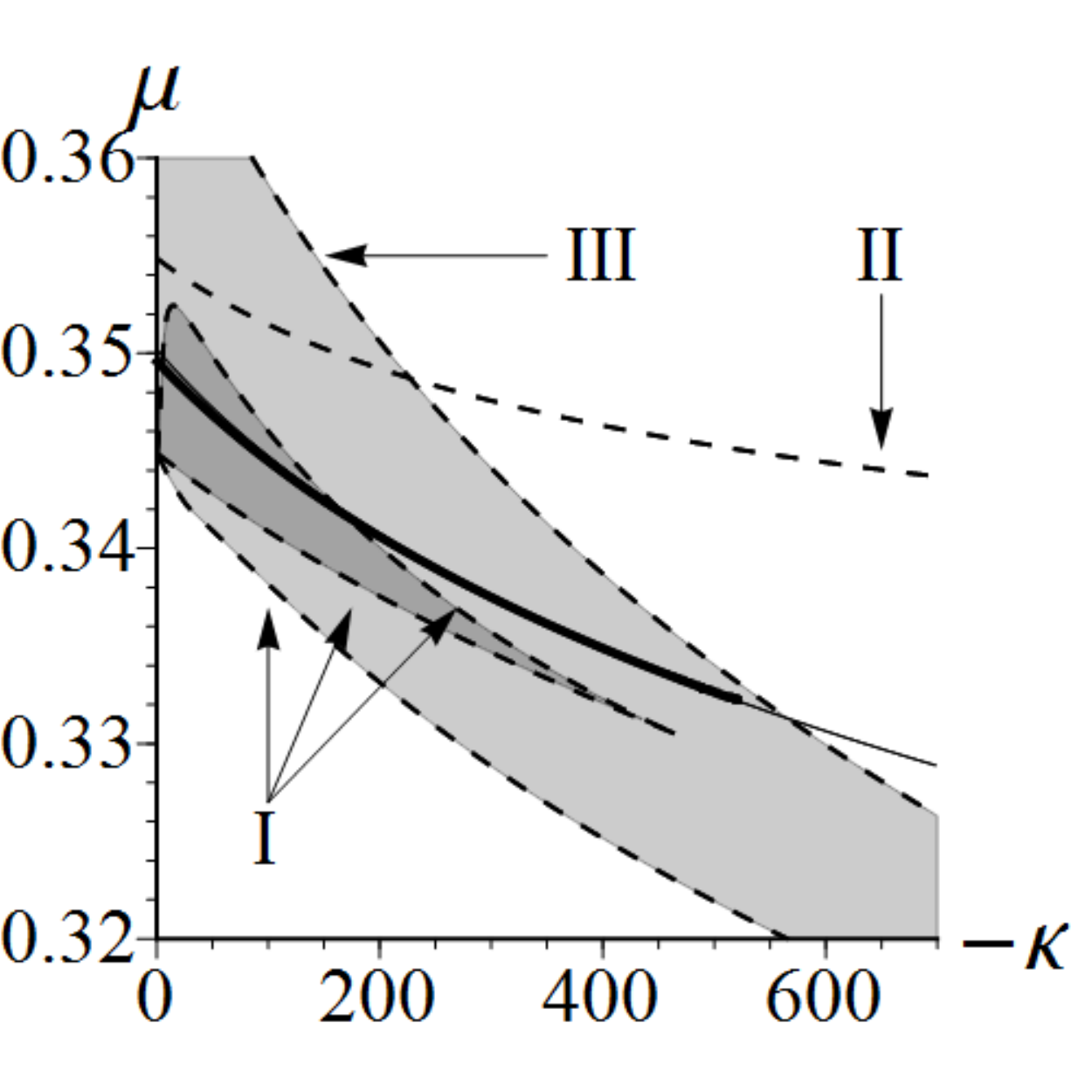}}
\subfigure[]{\label{grafAuxPotQuimSolNJLHCWtau14chilim}\includegraphics[width=0.24\textwidth]{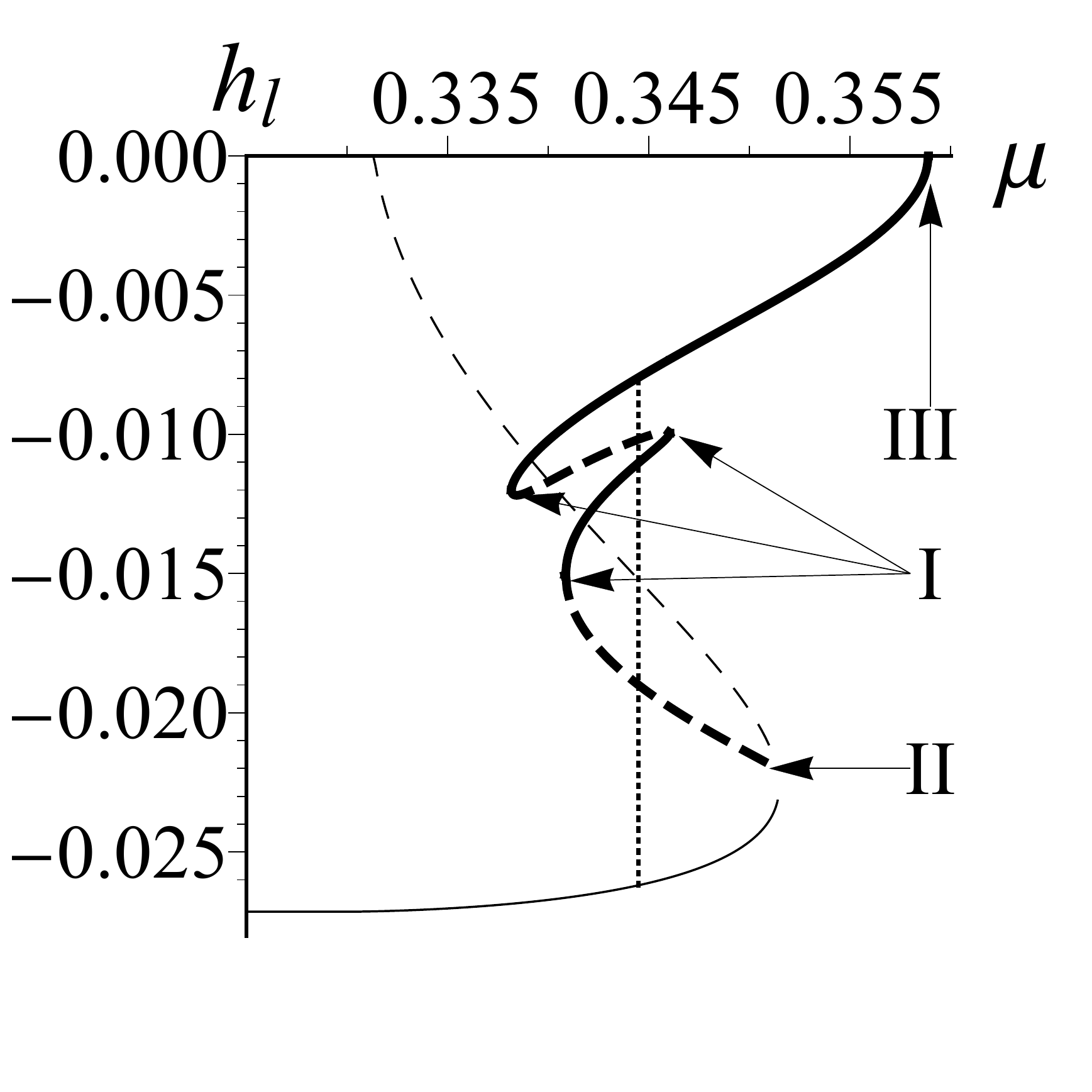}}
\caption{Critical chemical potentials for the case with flavor mixing in the
chiral limit and $\tau=1.4$. Their meaning is made evident in the right-hand side
panel where the light chiral condensate is shown as a function of the chemical
potential for $\kappa=-100~\mathrm{GeV}^{-5}$ case. The shaded area indicates
the existence of stable solutions with finite $q$ (light-grey: one solution, dark-grey: two solutions). 
A full line indicates the first order jump.}
\label{muCritVsKchilim}
\end{center}
\end{figure*}

Now let us consider the effect of flavor mixing with a massless strange quark
($m_u=m_d=m_s=0$). For values of the 't Hooft coupling constant lower than the
critical value ($-\kappa<-\kappa_{crit}=467\mathrm{GeV}^{-5}$) this flavor
mixing gives rise to the appearance of two new solution branches (these, when
shifted to higher chemical potentials due to the inclusion of a finite current
mass, give rise to the shark fin-like structure described in the previous
section). One of these is locally stable but the global minimum still
corresponds to the solution with a lower chiral condensate and a larger value of $q$ (see
Fig.~\ref{SolNJLHtau14K100chilim} for the  $\kappa=-100~\mathrm{GeV}^{-5}$
example). The first order jump, which is indicated by the full line in Fig.~\ref{muCritVsKchilim}, goes to a finite-$q$ solution for
$-\kappa<521~\mathrm{GeV}^{-5}$, and to the trivial solution (vanishing chiral
condensates) for stronger flavor mixing. As before, there is an additional
solution branch starting at a much higher chemical 
potential with finite but asymptotically vanishing $h_i$ and $q$.

\subsection{Comparison to parameter sets}

In an attempt to keep the discussion in the most general terms possible, we have up
to this point considered very few restrictions on the model parametrization. In
Table~\ref{parametersets} we list the values of the parameter $\tau$ extracted from several
sets of parameters used in the literature (when needed, a conversion of the cutoff from the 3D case to its equivalent
covariant value was done), used to fit the pseudoscalar spectrum given 
in Table~\ref{mesonspetra} for several variants of the 3-flavor NJL model. In sets (a-c)
isospin breaking was also considered, hence the values indicated in the table are the
averages over the isospin multiplets. In set (c) the mixing angle is not indicated in the respective paper.
Except for set (a), one observes a spread in the values of $\tau$ for the
different sets comprised between $1<\tau<1.5$, and a mixing interaction strength
$600 <-\kappa < 1800$ GeV$^{-5}$.

Despite the difference in the values for 
$\tau$,
%the curvature, 
from the point of view
of the value of $\kappa$, sets (c,e,f,g) should best fit to the case
$\kappa=-1000~\mathrm{GeV}^{-5}$ described above, while set (b) is eventually
best accomodated with $\kappa=-1800~\mathrm{GeV}^{-5}$, and set (d) by the
$\kappa=-500~\mathrm{GeV}^{-5}$. From the two-flavor case one can deduce that
diminishing $\tau$  corresponds to widening of the gap between the branch close to
the $\mu \sim {\hat M}$ and the one appearing at larger values of $\mu$.     

%%%%%%%%%%%%%%%%%%%%%%%%%%%%  TABLE 1  %%%%%%%%%%%%%%%%%%%%%%%%%%%%%%%%%%%%
%\vspace{0.5cm}
%\noindent
\begin{table*}
\caption{Parameter sets of the model: $\hat m, m_s$, and $\Lambda$ are given
         in MeV. The couplings have the following units: $[G]=$ GeV$^{-2}$,
         $[\kappa ]=$ GeV$^{-5}$. We also show
         the values of constituent quark masses $\hat M$ and $M_s$ in MeV.
         }
\label{parametersets}
\begin{tabular*}{\textwidth}{@{\extracolsep{\fill}}lrrrrrrrrl@{}}
\hline
Sets & \multicolumn{1}{c}{$\hat m$}
     & \multicolumn{1}{c}{$m_s$}
     & \multicolumn{1}{c}{$\hat M$}
     & \multicolumn{1}{c}{$M_s$}
     & \multicolumn{1}{c}{$\Lambda$}
     & \multicolumn{1}{c}{$G$}
     & \multicolumn{1}{c}{$-\kappa$}
     & \multicolumn{1}{c}{$\tau$}
     & \multicolumn{1}{c}{Ref.}  \\
\hline
a  & 5.5 & 135.7 & 335 & 527 & 1263  & 9.21  & 121   & 2.23 &
\cite{Hatsuda:1994pi} \\
b  & 7.7 & 159 & 315 & 486 & 805  & 11.58  & 1775   & 1.14 &
\cite{Reinhardt:1988xu}\\
c  & 7.7 & 159 & 315 & 508 & 805  &12.6  & 1183   & 1.24 &
\cite{Reinhardt:1988xu} \\
d  & 5.3 & 170 & 315 & 513 & 920  &8.98  & 687   & 1.14  & \cite{Osipov:2004mn} 
\\
e  & 6.1 & 185 & 380 & 576 & 830  &12.6  & 1116   & 1.32  & \cite{Osipov:2004mn}
 \\
f  & 5.8 & 183 & 348 & 544 & 864  &10.8  & 921   & 1.23  & \cite{Osipov:2007mk} 
\\
g  & 6.3 & 194 & 398 & 588 & 820  &13.5  & 1300   & 1.38  & \cite{Osipov:2006xa}
 \\
\hline
\end{tabular*}
\end{table*}
%\vspace{0.2cm}
%%%%%%%%%%%%%%%%%%%%%%%%%%%%%%%%%%%%%%%%%%%%%%%%%%%%%%%%%%%%%%%%%%%%%%%%%%%
%%%%%%%%%%%%%%%%%%%%%%%%%%%%  TABLE 2  %%%%%%%%%%%%%%%%%%%%%%%%%%%%%%%%%%%%
%\vspace{0.1}
%\noindent
\begin{table*}
\caption{The pseudoscalar  masses, weak decay constans and condensates  (in MeV)
for the
different sets. the mixing angle $\theta$ is in degrees.}
\label{mesonspetra}
\begin{tabular*}{\textwidth}{@{\extracolsep{\fill}}lrrrrrrrrl@{}}
\hline
Sets     & \multicolumn{1}{c}{$m_\pi$}
     & \multicolumn{1}{c}{$m_K$}
     & \multicolumn{1}{c}{$m_\eta$}
     & \multicolumn{1}{c}{$m_{\eta'}$}
     & \multicolumn{1}{c}{$f_\pi$}
     & \multicolumn{1}{c}{$f_K$}
     & \multicolumn{1}{c}{$-<{\bar u} u>^{\frac{1}{3}}$}
     & \multicolumn{1}{c}{$-<{\bar s} s>^{\frac{1}{3}}$}
     & \multicolumn{1}{c}{$\theta$}
\\
\hline
a     & 138 & 495.7 & 487. & 957.5  & 93  & 97.7 & 245 & 191 & -21 \\ 
b     & 140 & 495 & 502. & 1244  & 93.3  & 97.7 & 218 & 178 & -31 \\
c     & 140 & 495 & 441. & 958.  & 93.3  & 100.2 & 218 & 166 & \\
d     & 138 & 494 & 487. & 958.  & 92.  & 121. & 244 & 204 & -12\\
e     & 138 & 499 & 477. & 958.  & 92.  & 115.8& 233 & 182 & -15\\
f     & 138 & 494 & 476. & 986.  & 92.  & 118. & 237 & 191 &-13.6 \\
g     & 138 & 494 & 476. & 986.  & 92.  & 114.& 229 & 172 & -14 \\ 
\hline
\end{tabular*}
\end{table*}
%%%%%%%%%%%%%%%%%%%%%%%%%%%%%%%%%%%%%%%%%%%%%%%%%%%%%%%%%%%%%%%%%%%%%%%%%%%
\vspace{0.5cm}

\section{Conclusions}
We have made a thorough analysis of the instability of the strongly-interacting three-flavor quark matter at zero temperature with respect to the
formation of a spatially modulated inhomogeneous phase. 
%in a strongly interactingsystem of up, down and strange quarks 
We have used the NJL model augmented with the 't Hooft interations.

In the case with no flavor mixing there is a range of the model parameter $\tau$,
where we obtain a finite chemical potential window with energetically
favorable inhomogeneous solutions. There is a critical chemical
potential above which the favored solution is always an inhomogeneous one. Above
a critical value of $\tau$ these windows merge. This behavior has already been
reported (albeit with a different regularization procedure) in \cite{Carignano:2011gr}.
%where the case is made against this being a regularization effect.  

The inclusion of flavor mixing with the strange quark of a physical current mass
introduces new features into the model. For a fixed value of $\tau$ and depending on the
strength of the flavor mixing term we can have several scenarios. One of
these is the existence of an additional finite window for an inhomogeneous phase,
starting in a second order transition and ending in a first order transition
(meaning there exist two first order transitions). These two chemical
potential windows can also be connected, resulting in an interval (delimited by
two first transitions), where the dynamical mass behaves non-monotonically.
These features appear as a result of the shifting to higher chemical potential
of the new solutions induced by flavor mixing due to the presence of a
finite strange current mass.

There is a rich structure of solutions induced by the flavor mixing of the
strange quark with the spatially modulated light-quark sector. The main highlights
are:

\begin{itemize}
\item A new phase of globally stable inhomogeneous solutions ($q\ne0$) emerges,
covering a chemical potential  interval of several tens of MeV, for chemical
potential values below and in the neighborhood of  the mass of the vacuum value of the strange quark
mass. This phase occurs for physical strange current quark masses, as
well as in the ${\rm SU}_3$ chiral limit, in a wide range of the values of the 't Hooft
determinant strength $\kappa$.
\item Depending on the value of $\kappa$, this phase is either separated from the set of
inhomogeneous solutions known to occur at chemical potentials close to the
constituent quark mass of the light quarks in the vacuum, or joins with it. The interval
between these two branches of solutions shrinks with the increasing value of
$\kappa$ and there one only finds solutions with vanishing light quark condensates
with indeterminate value of $q$.
\item Besides these two branches, a third one exists at yet higher chemical
potential (larger than the strange quark mass). This type of solutions were
discussed in the literature in connection with the the ${\rm SU}_2$ case, and are
present in our three-flavor study as well.
\item Above a certain critical value of $\kappa$, there is a first order
transition which connects the solutions with different values of $q$, which prevail as being
the globally stable ones in the asymptotic regime $q\rightarrow 2 \mu$.
\item In the chiral limit, these branches of inhomogeneous solutions  have
overlapping chemical potential windows. Interestingly, one of the branches has in this case a
strange condensate which is lower (in absolute value) than the light condensate and is the
globally stable solution.
\end{itemize}

We conclude that flavor mixing acts as a catalyst for the emergence of globally
stable inhomogeneous solutions in zero-temperature quark matter. 
%as functions of the baryonic  chemical potential. 

\acknowledgments
This work has been supported by the Funda\c{c}\~ao para a Ci\^encia e Tecnologia, project: CERN/FP/116334/2010, developed under the iniciative QREN, financed by UE/FEDER through COMPETE - Programa Operacional Factores de Competitividade and the grant SFRH/BPD/63070/2009/. This research is part of the EU Research Infrastructure Integrating Activity Study of Strongly Interacting Matter (HadronPhysics3) under the 7th Framework Programme of EU, Grant Agreement No. 283286. W. Broniowski acknowledges the support of the Polish National Science Centre, grant DEC-2011/01/B/ST2/03915.

\bibliography{NJLCW}
\end{document}